\documentclass{aa}  
\usepackage{natbib}
\bibpunct{(}{)}{;}{a}{}{,}
\usepackage{graphicx}
\usepackage{txfonts}
\defcitealias{2016A&A...588A..56B}{Paper\,I}
\begin{document} 
\title{
Short-term variability and mass loss in Be stars
}
\subtitle
{
III. BRITE and SMEI satellite photometry of 28\,Cygni 
\thanks
{
Based in part on data collected by the BRITE-Constellation satellite mission,
built, launched and operated thanks to support from the Austrian
Aeronautics and Space Agency and the University of Vienna, the
Canadian Space Agency (CSA), and the Foundation for Polish Science \&
Technology (FNiTP MNiSW) and National Science Centre (NCN). 
}
}
\author{D.\,Baade\inst{1}
\and
A.\,Pigulski\inst{2}
\and 
Th.\,Rivinius\inst{3}
\and
A.C.\,Carciofi\inst{4}
\and
D. Panoglou\inst{5}
\and
M.\,Ghoreyshi \inst{4, 6}
\and
G.\,Handler\inst{7}
\and
R.\,Kuschnig\inst{8, 9}
\and
A.F.J.\,Moffat\inst{10}
\and
H.\,Pablo\inst{10}
\and
A.\,Popowicz\inst{11}
\and
G.A.\,Wade\inst{12}
\and
W.W.\,Weiss\inst{9}
\and 
K.\,Zwintz\inst{13}
}

\institute
{
European Organisation for Astronomical Research in the 
Southern Hemisphere (ESO), Karl-Schwarzschild-Str.\,2, 
\newline
85748 Garching b.\ M\"unchen, Germany;   
\email{dbaade@eso.org}
\and
Astronomical Institute, Wroc{\l}aw University, Kopernika 11, 
51-622 Wroc{\l}aw, Poland
\and
European Organisation for Astronomical Research in the 
Southern Hemisphere (ESO), Casilla 19001, Santiago 19, Chile 
\and
Instituto de Astronomia, Geof\'{\i}sica e Ci\^encias Atmosf\'ericas, Universidade de S\~ao Paulo, \\Rua do Mat\~ao 1226, Cidade Universit\'aria,
05508-900 S\~ao Paulo, SP, Brazil\
\and
Observat\'orio Nacional, Rua General Jos\'e Cristino 77, S\~ao
Crist\'ov\~ao RJ-20921-400, Rio de Janeiro, Brazil
\and 
Research Institute for Astronomy and Astrophysics of Maragha (RIAAM), 
Maragha, P.O.\ Box 55134-441, Iran
\and
Nicolaus Copernicus Astronomical Center, ul.\,Bartycka 18, 00-716 
Warsaw, Poland
\and
Institut für Kommunnikationsnetze und Satellitenkommunikation, Technical 
University Graz, Inffeldgasse 12, 8010 Graz, Austria
\and
Institute of Astrophysics, University of Vienna, Universit{\" a}tsring 1, 
1010 Vienna, Austria
\and
D{\' e}partement de physique and Centre de Recherche en Astrophysique du 
Qu{\' e}bec (CRAQ), Universit{\' e} de Montr{\' e}al, C.P. 6128, 
Succ.\,Centre-Ville, Montr{\' e}al, Qu{\' e}bec, H3C 3J7, Canada 
\and
Institute of Automatic Control, Silesian University of Technology,  
Akademicka 16, Gliwice, Poland
\and
Department of Physics, Royal Military College of Canada, PO
Box 17000, Stn Forces, Kingston, Ontario K7K 7B4, Canada
\and
Universit{\" a}t Innsbruck, Institut f{\" u}r Astro- und Teilchenphysik, 
Technikerstrasse 25, 6020 Innsbruck, Austria
}

\date{Received:  ; accepted:  }
 
\abstract 
{Be stars are important reference laboratories for the investigation of
viscous Keplerian discs.  In some cases, the disc feeder mechanism involves
a combination of nonradial pulsation (NRP) modes.

}
{Can high-cadence photometry shed further light on the role of NRP modes 
in facilitating rotation-supported mass loss?

} 
{The BRITE Constellation of nanosatellites obtained mmag photometry of
  28\,Cygni for 11 months in 2014-2016.  Observations with the Solar
  Mass Ejection Imager (SMEI) in 2003-2010 and 118 H$\alpha$ line
  profiles, half of them from 2016, were added.

} 
{For decades, 28\,Cyg has exhibited four large-amplitude 
  frequencies:
  two closely spaced frequencies of spectroscopically confirmed $g$
  modes near 1.5\,c/d, one slightly lower exophotospheric ({\v S}tefl)
  frequency, and at 0.05\,c/d the difference ($\Delta$) frequency
  between the two $g$ modes.  This top-level framework is
  indistinguishable from $\eta$ Cen (Paper I), which is also very
  similar in spectral type, rotation rate, and viewing angle.  The
  circumstellar ({\v S}tefl) frequency is the only one that does
  not seem to be affected by the $\Delta$ frequency.  The amplitude of
  the $\Delta$ frequency undergoes large variations; around 
  maximum the amount of near-circumstellar matter is increased, and
  the amplitude of the {\v S}tefl frequency grows by some factor.
  During such brightenings dozens of transient spikes appear in the
  frequency spectrum, concentrated in three groups.  Only eleven
  frequencies were common to all years of BRITE observations.

} 
{Be stars seem to be controlled by several coupled clocks, most of
  which are not very regular on timescales of weeks to months but
  function for decades. The combination of $g$ modes to the slow
  $\Delta$ variability and/or the atmospheric response to it appears
  significantly nonlinear.  Like in $\eta$\,Cen, the $\Delta$
  variability seems the main responsible for the modulation of the
  star-to-disc mass transfer in 28\, Cyg.  A hierarchical set of
  $\Delta$ frequencies may reach the longest timescales known of the
  Be phenomenon.

}

\keywords{
Circumstellar matter -- Stars: emission line, Be -- Stars: mass loss
-- Stars: oscillations -- Stars: individual: \object{28
Cyg}
}

\titlerunning{Satellite photometry of Be star 28\,Cygni}
\authorrunning{D.\,Baade et al.}

   \maketitle
%

\section{Introduction} 
\subsection{Be stars}

Be stars owe their designation to the occurrence of emission lines
forming in a self-ejected Keplerian disc.  The latest broad review of
the physics of Be stars is available from \citet{2013A&ARv..21...69R}.
This so-called Be phenomenon poses two core challenges:

\noindent
{\it a) What enables Be stars to toss up sizeable amounts of mass?}
Next to rapid rotation \citep{2005A&A...440..305F,
  2012A&A...538A.110M}, nonradial pulsations (NRPs) seem to be the
most common property known of Be stars \citep[][see also their Table 1
for pulsations of individual stars]{2013A&ARv..21...69R}.  Following
earlier suspicions, long series of echelle spectra of \object{$\mu$
  Cen} finally enabled \citet{1998ASPC..135..343R} to demonstrate that
mass-loss events in Be stars can be triggered by the temporary
combination of several NRP modes.  Paper\,I in this series
\citep{2016A&A...588A..56B} reported the second such example, based on
space photometry with BRITE-Constellation.  Also in some other Be
stars, rotation-assisted multi-mode NRP seems a sufficient condition
for the occurrence of discrete or repetitive mass-loss events \citep{
  2016arXiv161101113B, 2017ASPC..508...93B}.  Spectroscopically
strictly single-periodic stars like \object{28\,$\omega$\,CMa}
\citep{2003A&A...411..167S} with large-amplitude disc variability
caution that the condition of multiple modes may not be necessary.
But space photometry of \object{28\,$\omega$\,CMa} conveys a much more
complex picture \citep{2016arXiv161101113B}.  Models for NRP-driven
mass loss from rapidly rotating stars
\citep[e.g.,][]{1986A&A...163...97A, 2016ASPC..506...47K} do not yet
take into account strong interactions between a small number of modes.
Magnetic toy models are faced with  
non-detections \citep{2016ASPC..506..207W}.

\noindent
{\it b) What governs the interface between star and disc?}  The
interplay between mass injection and viscosity rules the life cycle of
Be-star discs. The main effect of the viscosity is a redistribution of
the specific angular momentum of the ejecta.  Only a small fraction of
the gas attains Keplerian velocities while the rest falls back to the
star.  During phases of replenishment, a Keplerian disc can form
\citep{1991MNRAS.250..432L}.  When the supply of new gas is shut off,
the inner disc quickly switches from decretion to accretion.
During the further course of time, this transition zone slowly grows,
and eventually most of the disc becomes an accretion disc
\citep{2012ApJ...756..156H, 2012ApJ...744L..15C}.  The disc is cleared
from the inside out as also observations suggest
\citep{2001A&A...379..257R}.  This somewhat violent coalescence of
mass ejection and reaccretion is probably not further complicated by
a significant wind of immediate stellar origin
\citep{1989MNRAS.241..721P, 2014A&A...564A..70K}.  Instead of directly
from the star, winds in Be stars may obtain a fair part of
their supply of matter through radiative ablation of the disc
\citep{2013A&ARv..21...69R, 2016MNRAS.458.2323K}.

Near-stellar matter can be approximated as a pseudo-photosphere
\citep{1983HvaOB...7...55H, 2015MNRAS.454.2107V} that reemits stellar
radiation.  At optical wavelengths, the region sampled is within very
few radii of the stellar photosphere.  The observable signature
depends on the viewing angle \citep{2012ApJ...756..156H} and is
largest for near face-on orientations.  Circumstellar gas in the line
of sight absorbs light (see Fig.\,\ref{VtoR} for some illustration 
of the aspect-angle dependency).  Precision photometry can extract much of the
information encoded in the short- and mid-term variability of Be stars
about both the central star and the circumstellar disc.  It requires
careful distinction between genuine pulsations and more volatile
frequencies that may serve as diagnostics of circumstellar gas
dynamics; see \citetalias{2016A&A...588A..56B} and \citet[][Paper
II]{2016A&A...593A.106R}.  A prominent part of the latter seem to be
the so-called \citepalias{2016A&A...588A..56B} {\v S}tefl frequencies
\citep[][Sect.\,\ref{circumstellar}]{1998ASPC..135..348S}.  For
unknown reasons they are roughly 10\% lower than
the dominant spectroscopic frequency.

The relevance of Be stars also has dual, stellar and circumstellar,
roots.  Be stars combine near-critical rotation with pulsation, making
them a testbed of evolution models, and the optical angular diameters
of Be discs are the largest known of any viscous (accretion or
decretion) discs.  For instance, quasar accretion discs measure a few
lightdays across in microlensing events
\citep[e.g.,][]{2012ApJ...751..106J}.  AGN in Seyfert galaxies are
less powerful so that 0.01\ pc is an upper limit for them.  The most
nearby Seyfert galaxies such as the Circinus Galaxy, NGC\,1068, and
NGC\,4725 are at 5-20\,Mpc \citep{2010ApJ...725.2270P}, bringing the
angular sizes of their AGN to about 0.2\,mas.

Most accreting stars are compact objects with small accretion discs.
Only systems with high mass-exchange rates and consisting of two
main-sequence (MS) stars or an MS star and a giant develop sizeable
discs.  Such discs have diameters of no more than a few times that of
the MS mass gainers, which typically are cooler and smaller than
early-type Be stars.  For instance, \citet{2008ApJ...684L..95Z}
measured the dimensions of the accretion disk of $\beta$ Lyrae in the
$H$ band as 1 $\times$ 0.6\,mas.  In their Table\,2,
\citet{2013A&ARv..21...69R} compiled diameters from optical
long-baseline interferometry at different wavelengths of 22 Be stars.
They span approximate ranges of 0.5-4\,mas and 1.5-15 stellar radii.
In addition to being more extended, Be discs are also quite bright.

These properties permit accretion-disc physics to be studied at high
resolution in space, velocity, and time and at excellent
signal-to-noise ratio.  For the realisation of this diagnostic value,
advanced radiative transfer tools have been developed: e.g., the
radiatively-driven-wind model SIMECA \citep[][and references
therein]{2001A&A...367..532S}, HDUST \citep{2006ApJ...639.1081C}, and
BEDISK \citep{2007ApJ...668..481S}.  HDUST has also been combined with
a 1D hydrodynamics grid code \citep{2007ASPC..361..230O} and a 3D
smoothed particle hydrodynamics code \citep{2002MNRAS.337..967O} to
form a Viscous Decretion Disc (VDD) model as employed by, e.g.,
\citet{2012ApJ...756..156H} and \citet{2017arXiv170406751P}.

Some of the diagnostically most valuable observations of Be stars are
delivered by space-borne photometers.  MOST
\citep{2005ApJ...635L..77W}, CoRoT \citep{2009A&A...506...95H,
  2012A&A...546A..47N}, and Kepler \citep{2015MNRAS.450.3015K} have
pioneered the field, and the Solar Mass Ejection Imager
\citep[SMEI,][]{2004SoPh..225..177J, 2013SSRv..180....1H} produced
stellar photometry as a byproduct for nearly eight years.  For an
overview see \citet{2016arXiv161007188R}.  Currently, BRITE
Constellation (Sect.\,\ref{BRITEobs}) is accumulating a large database
of naked-eye stars.  A first cursory preview of the broad range of
variabilities observed by BRITE in Be stars was already published
\citep{2016arXiv161101113B, 2017ASPC..508...93B}.  Similarly to
\citetalias{2016A&A...588A..56B}, the present work uses BRITE space
photometry to study the variability of a well-known single object.

\subsection{28\,Cygni} 
\label{28cygintro} 

28\,Cyg (= HR\,7708 = HD\,191610 = HIP\,99303) is a seemingly single
B2 IV(e) star \citep{1982ApJS...50...55S} with
$v \sin i = 320$\,km/s \citep{2003A&A...411..229R} and no record of
shell absorption lines.  \citet{2006A&A...459..137R} list six shell
stars between B1 and B3 and an average $v$\,sin\,$i$ of 335\,km/s on
the Slettebak scale \citep{1982ApJS...50...55S} with a maximum of
400\,km/s.  For a disc with opening angle $\leq20^\circ$, the
inclination angle, $i$, of 28\,Cyg should, therefore, not exceed
$75^\circ$.  Comparison to the sample of \citet{2003A&A...411..229R},
which does not include shell stars, suggests a lower limit around
$40^\circ$.  The VDD model predicts \citep{2012ApJ...756..156H}
anticorrelations between colour indices $U-B$ and $B-V$ as observed by
\citet{1997A&AS..125...75P} in 28\,Cyg as well as between $V$ and $B-V$.
A quantitative comparison is impossible because the calculations use a
no-disc state for reference, which probably did not occur in 28\,Cyg,
and some maximum state, which the observations cannot be related to.
The observed $U-B$ vs.\ $B-V$ slope is compatible with
$30^\circ \leq i \leq 70^\circ$ (\citet{2012ApJ...756..156H} do not
illustrate intermediate aspect angles).

\begin{figure*}
\includegraphics[width=11.8cm,angle=-90]{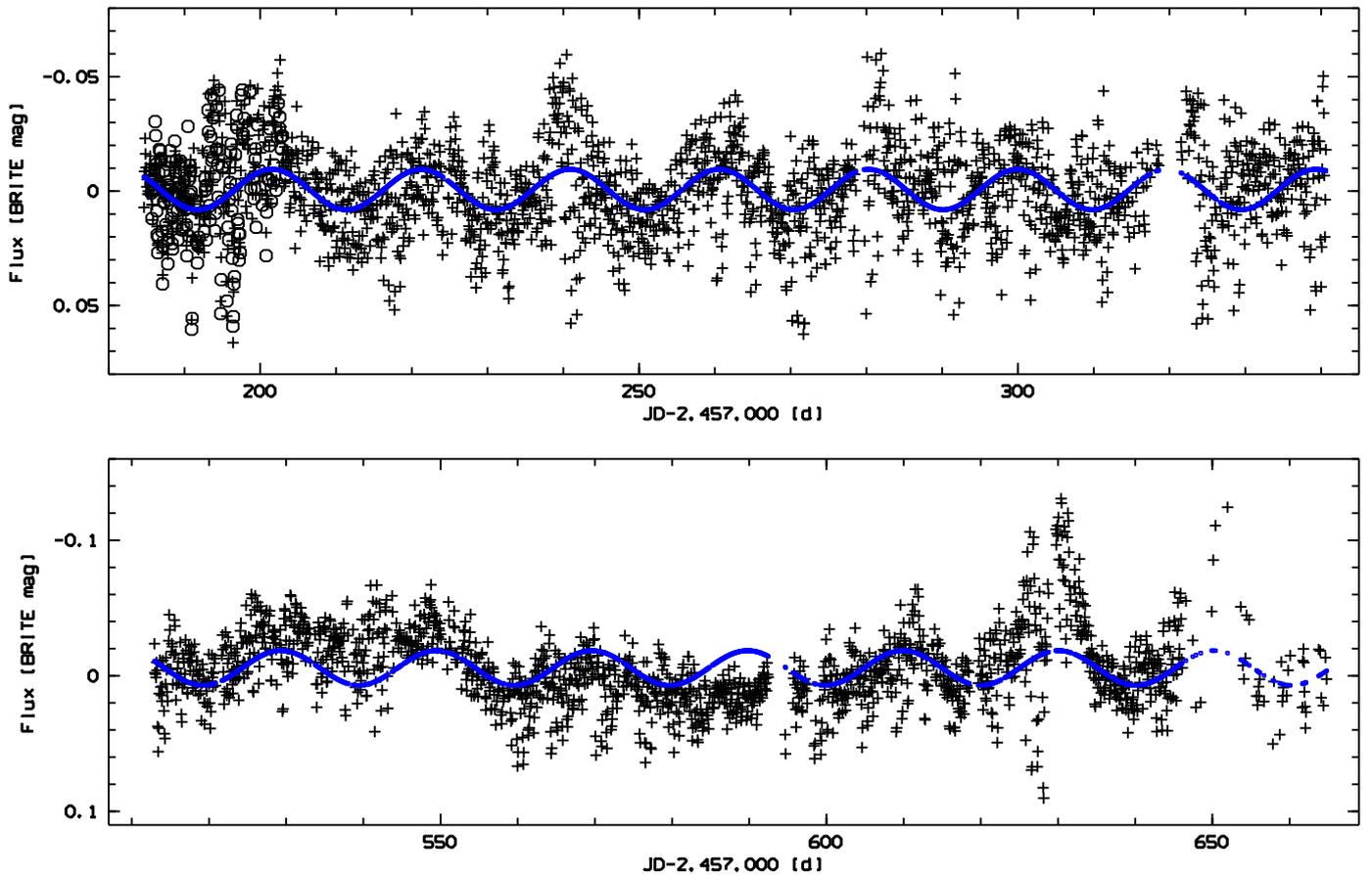}
\caption{
Light curve of 28\,Cyg observed with BTr ($+$) and UBr ($\circ$) in 2015 
(top) and BTr in 2016 (bottom).  
Fitted constant-amplitude sine curves with frequencies 0.0508\,c/d (2015) 
and 0.0496\,c/d (2016) are overplotted in blue and trace 
the respective $f_{\Delta32}$ ($\approx f_3 - f_2$). JD = Julian Date. 
The zero points of the magnitude scales are arbitrary.  Note the difference 
in magnitude scale between 2015 and 2016.  
Fifty-seven BeSS spectra were obtained between mJD\,632 and 733 (see 
Figs.\,\ref{BeSSperspHalpha} and \ref{VtoR}).
} 
\label{28LCf4} 
\end{figure*}

\begin{table*}
\caption{Overview of BRITE observations}
\label{overviewtab}
\centering
\begin{tabular}{l c c c r r c r r}
\hline\hline
Satellite name      & Orbital period & Year & Contig.\ time &  JD start  &  JD end &  Range in CCDT        & No.\ of 1-s & No.\ of TSA \\
(acronym)           & [min]          &      &  [min]        &  \multicolumn{2}{c}{-2,450,000} &  [\degr C] & exposures   & data points \\
\hline
BRITE-Lem           & 99.7           & 2014 &   7 - 10      &  6852      & 6867    &  13.7 - 22.5          &   3059      &  111        \\
(BLb)               &                & 2015 &  10 - 15      &  7274      & 7287    &  28.7 - 38.8          &   1781      &   41        \\
                    &                & 2016 &   9 - 14      &  7634      & 7636    &  25.2 - 26.6          &    727      &   20        \\
BRITE-Toronto       & 98.2           & 2014 &  10 - 16      &  6850      & 6890    &   7.7 - 20.3          &   9654      &  251        \\
(BTr)               &                & 2015 &   5 - 27      &  7185      & 7341    &   7.4 - 22.3          & 105675      & 2098        \\
                    &                & 2016 &   4 - 27      &  7513      & 7665    &   7.4 - 20.4          &  65454      & 1747        \\
UniBRITE            & 100.4          & 2015 &   7 - 16      &  7186      & 7204    &  27.2 - 29.9          &   7681      &  175        \\
(UBr)               &                &      &               &            &         &                       &             &             \\
\hline
\end{tabular}
\tablefoot{Suffixes 'b' and 'r' indicate the passband (blue/red).  'Contig.\ 
time' denotes the typical contiguous time interval per orbit 
during which exposures were made. CCDT is the temperature of the detector.  
The BTr observations after JD 2,456,877 
suffered from excess noise and reduced instrumental throughput and were 
discarded.}
\end{table*}

28\,Cyg was among the first Be stars in which short-term variability
was observed \citep{1977AJ.....82..353P, 1977AJ.....82..166G}.
Spectroscopy followed suit and detected periods near 0.64\,d
(frequency: 1.56\,c/d) \citep{1988ESASP.281b.117P,
  1990A&A...236..393P, 1994IAUS..162..100H} and 0.7\,d (1.4\,c/d)
\citep{1981PASP...93..460S, 1993A&A...269..343B}.
\citet{1997A&AS..125...75P} established a photometric peak-to-valley
(PTV) amplitude of about 0.1\,mag in $V$, and variations in $B-V$ and $U-B$
amounted to $\sim$0.07\,mag PTV in both colours without obvious
relation to the $V$ magnitude.  From a combination of a subset of these
data with other observations,
\citet{1994IAUS..162..102R1997A&AS..125...75P} inferred frequencies of
1.45\,c/d and 0.09\,c/d.  \citet{2001PASP..113..748P} found it
difficult to derive a single-periodic light curve and noted a
remarkably low variability of only a few 0.01\,mag on a timescale of
hundreds of days.  The spectroscopic studies presented signatures of
nonradial pulsation. Most intriguing are reports by
\citet{1988ESASP.281b.117P} and \cite{2000ASPC..214..375P} of
synchronous variations in photospheric and UV wind lines, i.e., a
hypothetical link between pulsations and mass loss (see
Sect.\,\ref{spectra}).

From 209 echelle spectra obtained in 1997 and 1998,
\citet{2000ASPC..214..232T} confirmed and refined the 1.56-c/d
frequency but also deduced a second one at 1.60\,c/d.  Both carried
the signature of low-order nonradial $g$-mode pulsation that is
typical of Be stars \citep{2003A&A...411..229R}.  Inspired by the
example of \object{$\mu$\,Cen} \citep{1998ASPC..135..343R}, the
authors searched for indications of enhanced mass loss repeating with
the beat frequency of 0.055\,c/d.  Only one outburst was clearly found
to coincide with a moment when both variations reached their maximal
amplitude.

\begin{figure} 
\includegraphics[width=6.6cm,angle=-90]{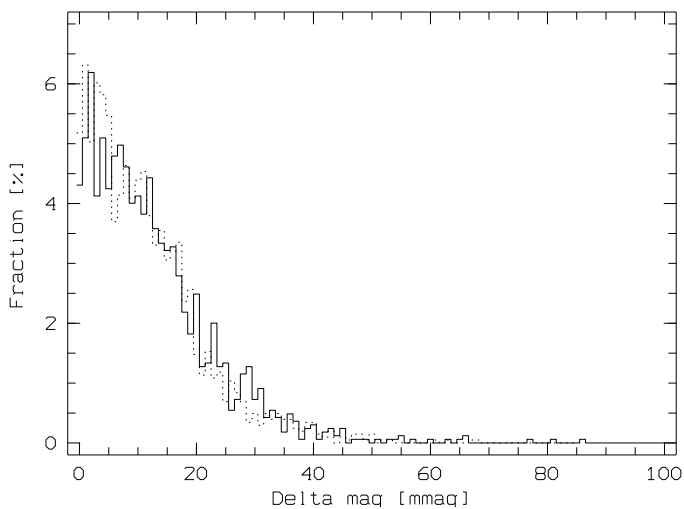}
\caption{Histogram of the magnitude differences between directly
  consecutive orbits (data points) of BRITE satellite BTr in 2015
  (dotted line) and 2016 (solid line). }
\label{deltaMagHistB} 
\end{figure}

\begin{figure} 
\includegraphics[width=3.6cm,angle=-90]{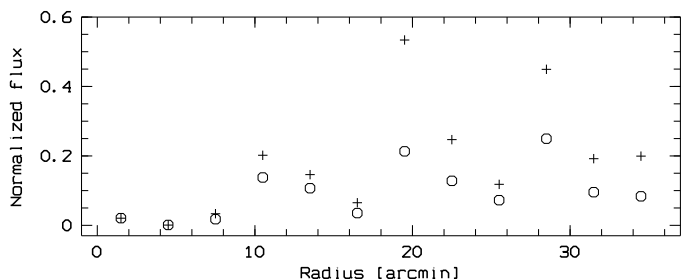}
\caption{$B$-band (crosses) and $V$-band (circles) fluxes (from
  SIMBAD) in annuli of 3\,arcmin width around \object{28\,Cyg}.  In
  both bands, the flux of 28\,Cyg was set to unity.  For BRITE
  observations in chopping (nodding) mode, the two positions of the
  point spread function (PSF) are contained in a window of
  11\,$\times$\,24\,arcmin$^2$.  The PSF of SMEI was elongated,
  $\sim$60\,arcmin wide, and not box-car shaped.  }
\label{SMEIfield} 
\end{figure}

\begin{figure}
\includegraphics[width=6.2cm,angle=-90]{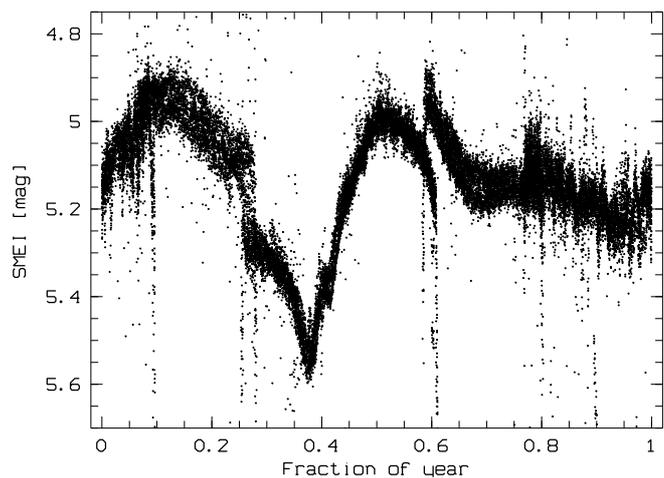}
\caption{
Mean annual SMEI light curve of 28\,Cyg (in SMEI magnitudes and clipped 
to the range shown). The bimodal distribution of values  during some 
phases is probably 
the result of imperfect cross-calibration of the cameras.  
Fraction 0.0 corresponds to 1 January.  
} 
\label{28lcSann} 
\end{figure}

\section{Observations, data reduction, and analysis methods}
\subsection{BRIght Target Explorer (BRITE)} 
\label{BRITEobs}

BRITE-Constellation consists of five nanosatellites as described by
\citet{2014PASP..126..573W}.  \citet{2016PASP..128l5001P} report on
pre-launch and in-orbit tests.  The pipeline processing from the raw
images to the instrumental magnitudes delivered to users is elaborated
by \citet{2017arXiv170509712P} who also report per-orbit photometric
errors of 2-6\,mmag for stars of comparable brightness as
\object{28\,Cyg}.  At $V$\,=\,4.9\,mag \citep{2002yCat.2237....0D}, the
star is nearly one magnitude brighter than the normal faint end of the
capabilities for low-amplitude variabilities with BRITE
\citep[][Kuschnig, in prep.]{2017arXiv170509712P}.  After a
proprietary period of typically one year, pipeline-processed BRITE
data are made public\footnote{URL:
  http://brite.craq-astro.ca/doku.php}.

The datasets for 28\,Cyg are summarized in Table\,\ref{overviewtab}.
BRITE-Constellation members Lem (BLb) and Toronto (BTr) observed
28\,Cyg for 15 and 40\,d, respectively, in 2014.  In a second visit in
2015, Lem and Toronto collected data for 13 and 156\,d, respectively,
and UniBRITE (UBr) contributed an additional 18 days.  BTr and BLb
revisited 28\,Cyg in 2016 for another 152 and 18\,d, respectively.  A
`b' (`r') in the abbreviation of a satellite refers to the blue
390-460\,nm (red 550-700\,nm) passband.  The observations in 2015 and
2016 were obtained in the so-called chopping mode, in which the
satellites nod between two positions slightly farther apart than the
width of the point spread function.  The difference between the on-
and off-position aperture photometry is less affected by the
progressing radiation damage of the detectors than stare-mode data are
\citep{2016PASP..128l5001P, 2016SPIE.9904E..1RP}.

The point-spread function (PSF) of BRITE varies strongly across the
field.  Its two nodding positions are enclosed in a window of
dimensions 11'\,$\times$\,24' \citep{2017arXiv170509712P}.  At any one
instant, the PSF covers about one-third of one nodding halve of the
window.  However, pointing and tracking errors introduce considerable
exposure-to-exposure jitter, thereby reducing the scope of background
removal.  For the flux contribution from other discrete sources in the
field see Fig.\,\ref{SMEIfield}.

The input data used are from data releases \citep{2017arXiv170509712P}
DR2 for the observations in 2014, DR3 for 2015, and DR5 in 2016.
Their preparation for time-series analysis (TSA) consisted of
detrending them for dependencies on position (X/Y) and detector
temperature.  The procedure was the same as in
\citetalias{2016A&A...588A..56B} except that a first very conservative
detrending for these three parameters of the data was already applied
to the 1-s exposure data.  A second such decorrelation was performed
with orbit-averaged data mostly with a view towards identifying and
eliminating obvious outliers and correcting for any residual trends.
For perfect detrending a priori knowledge about the intrinsic
variability is required.  Because this information does not exist, the
data were reduced several times to find the best possible compromise
between the removal (mostly by clipping) of artifacts and the
preservation of ephemeral events.  The detrending was applied
separately to each so-called setup (data strings obtained with the
same satellite attitude, etc.).  Before the merger of data from
different setups, small constant offsets were applied if shifts in
zeropoint seemed to require compensation.  The (pseudo-)Nyquist
frequency of the results is $f_{\rm orbit}/2 \approx 7.2$\,c/d.  The
resulting light curves are plotted in Fig.\,\ref{28LCf4}.

\begin{figure*}
\includegraphics[width=6.5cm,angle=-90]{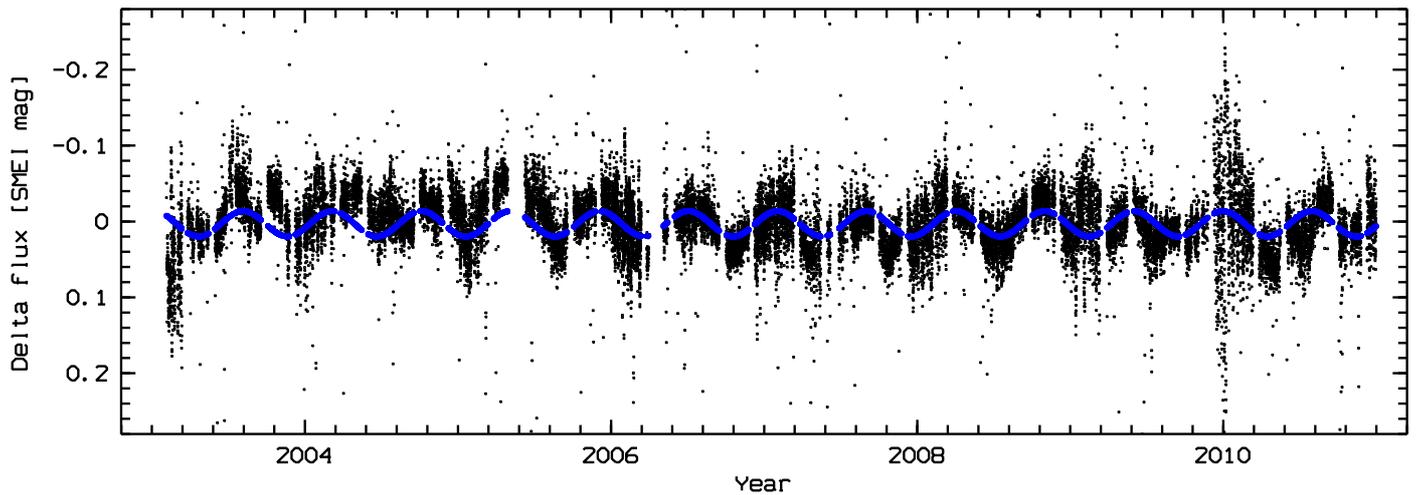}
\caption{
SMEI light curve of 28\,Cyg after subtraction of the instrumental annual 
variation shown in Fig.\,\ref{28lcSann} and after elimination of 
data from annual phases with elevated noise.  In addition, some 
extreme outliers are clipped in this figure.  No indication is visible 
of any photometric equivalent of the fading of the H$\alpha$ emission 
between 2005/2007 and 2013 (Sect.\,\ref{spectra}).  Overplotted in blue 
is a fitted sine curve with frequency 0.00471\,c/d.  
} 
\label{28lcJDS} 
\end{figure*}

Because BRITE does not perform differential photometry, it is not
really possible to determine the intrinsic accuracy beyond the
reference values provided by \citet{2017arXiv170509712P}.  Of course,
one could subtract all variabilities found by the TSA and, then,
compute some statistics.  However, if the purpose is to assess the
significance of the frequencies, this and similar methods lead to
circular reasoning.  As a substitute, Fig.\,\ref{deltaMagHistB} offers
the histogram of the magnitude differences between immediately
consecutive orbits of BTr in 2015 and in 2016 which produced the bulk
of the observations (see Table\,\ref{overviewtab}).  BTr has an
orbital frequency of 14.66\,c/d and a duty cycle of 5-27\%
(Table\,\ref{overviewtab}) so that for the observed multiple
frequencies of up to 3\,c/d (Sect.\,\ref{BRITEcommon}) the inter-orbit
variability is not negligible.  Therefore, Fig.\,\ref{deltaMagHistB}
only enables a worst-case estimate.

The TSA was performed after conversion of the fluxes to instrumental
magnitudes and subtraction of the mean.  Three datasets were fully
examined, namely the BTr observations from 2015 as well as 2016 and
their combination. The much smaller datasets from BLb are noisier and
only cover a fraction of the BTr time intervals so that they were
omitted from the analysis.  Because potentially disturbing
low-frequency variability was not seen, no further corrections were
attempted.

\subsection{Solar Mass Ejection Imager (SMEI)}
\label{SMEIobs}

The Solar Mass Ejection Imager \citep{2004SoPh..225..177J,
  2013SSRv..180....1H} was a secondary payload onboard the Coriolis
spacecraft and tasked with monitoring space weather in the inner solar
system.  The signal recorded for this purpose was the
Thomson-scattered solar light from ejected interplanetary electrons.
Coriolis was launched in 2003 January into an orbit with a period of
101.6 minutes.  The goal design lifetime was 5 years, but SMEI was
eventually powered off only in 2012 September.

SMEI delivered nearly 4$\pi$ surface brightness measurements with
three cameras with a field of view of $3 \times 120 $\,deg$^2$ each
and a basic angular sampling of 0.2$^\circ$ (0.1$^\circ$ for Camera
3).  An inner zone of avoidance of 20$^\circ$ radius enabled some
overlap between cameras.  Camera 3 observed the region closest to the
Sun, Camera 1 the zone farthest away from it.  Point sources drifted
through the combined field of view within $\sim$1.5\,min, and
continuous 4-s exposures were obtained with one frame-transfer CCD per
camera.  The core unfiltered photometric passband ranged from 450\,nm
to 950\,nm with a maximum near 700\,nm.  After an electronics failure
in 2006, the temperature of Camera 3 rose, and its performance
gradually degraded substantially.

The point spread function of the deliberately defocussed cameras
measures $\sim$$1^{\circ}$
across, is highly asymmetric, and varies strongly over the path along
which an object moved through the field of view
\citep{2007SPIE.6689E..0CH}.  With its aperture of 1.76\,cm$^2$,
SMEI could detect point sources of $\sim$$10^{\rm
  m}$.
Surface-brightness measurements with a precision of 0.1\% required
careful removal of stars brighter than $\sim$$6^{\rm
  m}$ from the images \citep{2007SPIE.6689E..0CH}.  In addition to
direct and scattered light from discrete solar-system objects, the
strongest background signal by far was the zodiacal light, which in
some directions exceeded the signal from solar mass ejections by two
orders of magnitude.  It was removed by careful modelling.  However,
apart from any systematic issues, the stellar SMEI data inherited much
photon noise from this background, to which the South Atlantic Anomaly
made additional contributions.  Radiation damage of the CCDs caused a
loss in sensitivity of 1.6\% per year and cosmetic problems.

Stellar fluxes \citep[in SMEI magnitudes;][]{2007SPIE.6689E..0BB}
extracted by the UCSD pipeline \citep{2005SPIE.5901..340H} are
available on the
Web\footnote{http:$//$smei.ucsd.edu/new\_smei/data\&images/stars/timeseries.html}.
The only accompanying telemetry are observing dates and times so that
it is not possible to analyze the observations separately for each
camera.  This has major repercussions as demonstrated below.  So many
years after the termination of the SMEI mission, the contact address
provided at the UCSD SMEI Web site seems no longer actively supported.
An example of observations with SMEI of relatively faint objects
(novae) is available from \citet{2010ApJ...724..480H} who provide
overall absolute errors around maximum (3.5-5.4\,mag) of
0.01-0.02\,mag.

\begin{figure} 
\includegraphics[width=6.6cm,angle=-90]{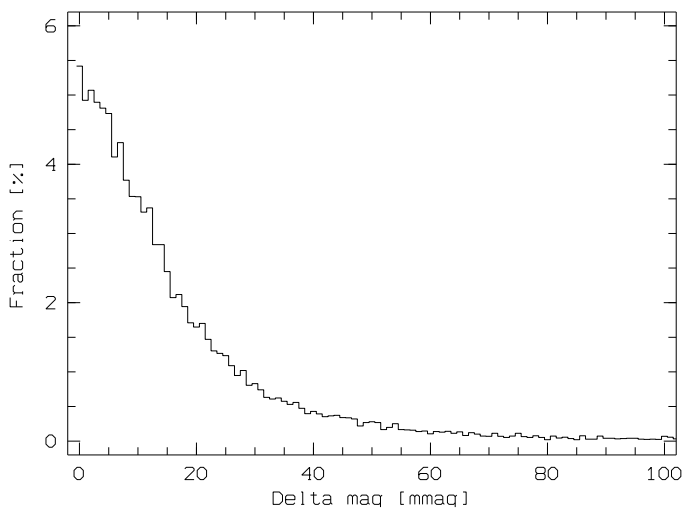}
\caption{Histogram of the magnitude differences between directly
  consecutive orbits (data points) of the SMEI data shown in 
  Fig.\,\ref{28lcJDS}.}
\label{deltaMagHistS} 
\end{figure}

SIMBAD \citep{2000A&AS..143....9W} lists 443 sources within
$0.6^\circ$ from \object{28\,Cyg}.  Only relatively few have
magnitudes in the $R$ band, the best match of the sensitivity of SMEI.
$B$ and $V$ fluxes for 323 and 314 sources, respectively in SIMBAD
account for 1.2 and 2.2 times as much flux as \object{28\,Cyg} alone
does.  The distribution of these fluxes in annuli of 3\,arcmin width
is shown in Fig.\,\ref{SMEIfield}.  

The SMEI observations of 28\,Cyg extended over 7.9 years from Feb.\,4,
2003 to Dec.\,30, 2010 so that the nominal frequency resolution is
$\leq$0.001\,c/d.  This dataset comprises a total of 34,335 data
points.  Since each data point results from one 101.6-minute orbit,
more than 80\% of all orbits contributed, and the nominal Nyquist
frequency is $\sim$7.1\,c/d.  After removal of extreme outliers, the
raw light curve (Fig.\,\ref{28lcSann}) revealed a strong need for
correction for instrumental annual and other long-term variability.
In addition, at some phases, the distribution of data points is
bimodal, which is possibly due to the combination of data from
different cameras.  After prewhitening for the annual variability by
subtracting means in phase intervals of 0.01, a considerable annual
pattern was still left because of season-dependent excess noise.
These annual phase intervals were removed from the dataset which
thereby shrunk to 26,157 data points.  Another 1,184 presumed outliers
were eliminated after its inspection.

\begin{figure*} 
\includegraphics[width=12.5cm,angle=-90]{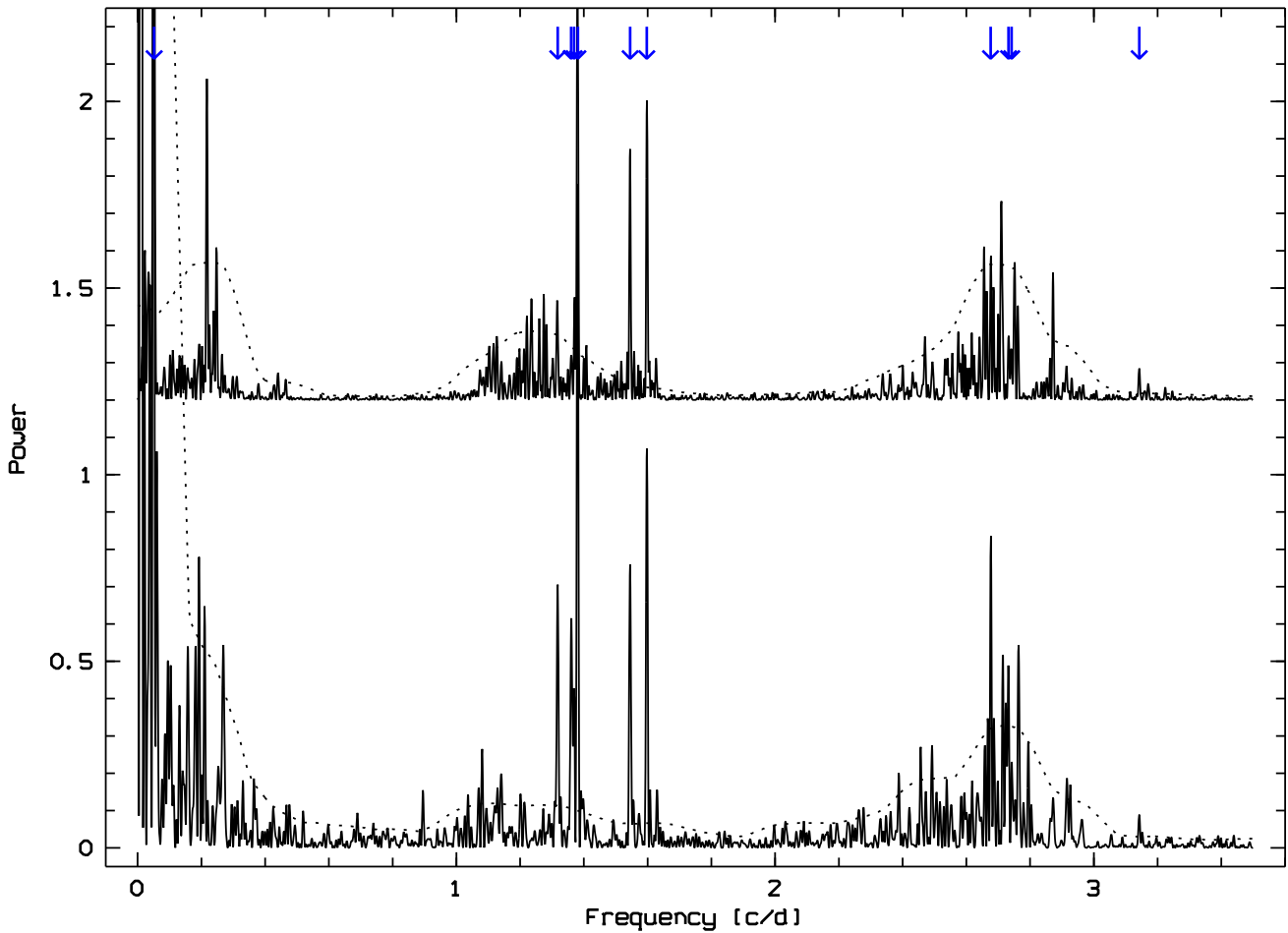}
\caption{ BRITE frequency spectrum (in arbitrary units) of 28\,Cyg.  Top:
  2015 (BTr and UBr), bottom: 2016 (BTr).  Arrows mark the frequencies
  included in Table \ref{28freq}).  The dashed lines represent the
  local sum of mean power and 3 $\times$ $\sigma$ (calculated after
  removal of the frequencies in Table\,\ref{28freq}).  Between
  3.5\,c/d and the nominal Nyquist frequency near 7.2\,c/d, there is
  virtually no power.  Frequency groupings occur at the approximate
  ranges 0.1-0.5\,c/d ($g_0$), 1.0-1.7\,c/d ($g_1$), and 2.2-3.0\,c/d
  ($g_2$)}.
\label{28powerB} 
\end{figure*}

\begin{figure*}
\includegraphics[width=18.6cm,angle=0]{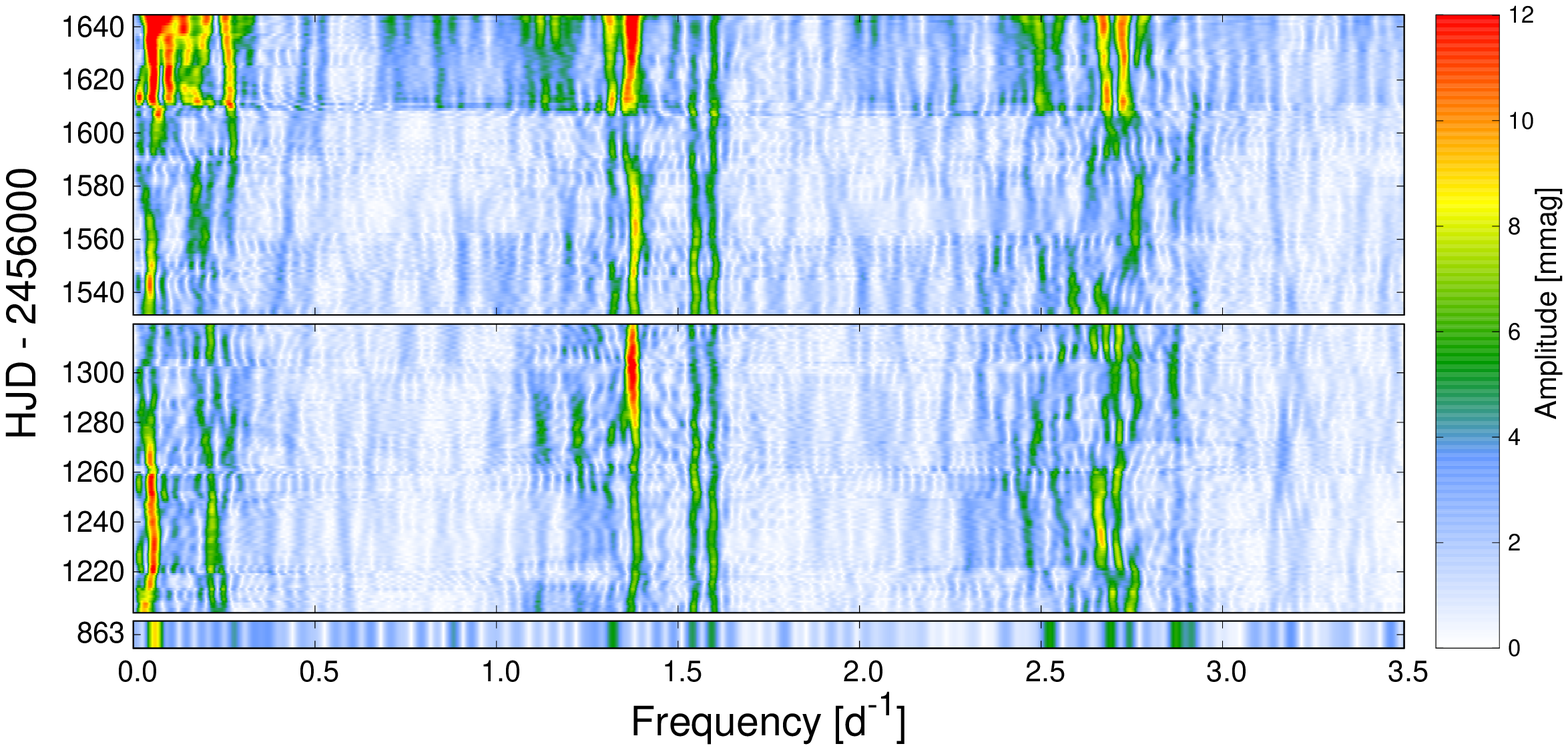}
\caption{ Time vs.\ frequency diagrammes for the BRITE red-filter
  photometry of 28\,Cyg in 2014 (bottom, a single time bin), 
  2015 (middle), and 2016 (top). The
  discrete-Fourier-transform frequency spectra were calculated in
  40-day intervals with a step of 1\,d. The common amplitude scale
  appears on the right.  Horizontal lines occur where data from
  different setups were stitched together. Note the diffuse appearance
  of $f_{\Delta32}$ (0.05\,c/d) and $f_1$ (1.38\,c/d, the {\v S}tefl frequency)
  as well as the cyclic anti-phased variability of
  the two $g$ modes $f_2$ (1.54\,c/d) and $f_3$ (1.60\,c/d).  }
\label{timefreq} 
\end{figure*}

Next, the spacing in time of the remaining 24,973 observations was
studied.  It can vary if the posted times correspond to the mean time
of the 4-s exposures co-added for each orbit and the number of the
latter varies because of rejected exposures.  It was hypothesized that
the best orbital data points are those that result from orbits without
rejected measurements and that they can be identified by their spacing
corresponding closely to the orbital period of 101.6 minutes.  This
selection reduced the number of data points to 23,708.  Before this
scheme was adopted, various others were explored, including
permutations of the above steps.  When the filtering according to the
time elapsed between consecutive data points was performed at the
beginning, it eliminated far more measurements than when done at the
end.  This demonstrates that the filtering of the annual light curve,
which contains some subjective elements, has an objective background.

In addition to the annual variability (Fig.\,\ref{28lcSann}), SMEI
data suffer instrumental variabilities faster than a year
but slower than those measured in the BRITE observations of
\object{28\,Cyg} (Sect.\,\ref{BRITEresults} and Table\,\ref{28freq}).
Figure\,\ref{28lcJDS} depicts the full light curve with all corrections
applied as described above.  For the observations with SMEI of
\object{$\alpha$ Eri} (B6\,Ve), \object{$\zeta$ Pup}, and
\object{$\zeta$ Oph} (O9.5\,V), \citet{2011MNRAS.411..162G},
\citet{2014MNRAS.445.2878H}, and \citet{2014MNRAS.440.1674H},
respectively, used a boxcar filter of width 10\,d to remove spurious
non-stellar signals.  In spite of the low accuracy of the individual
SMEI measurements, their large number and long time span permitted
these studies to detect significant variabilities at frequencies
0.7-7.2\,c/d with amplitudes down to a few mmag.  Experiments with
filter widths between 6\,d and 15\,d showed that in the range above
0.5\,c/d, where most of the BRITE frequencies occur, the results of
the TSA are not very sensitive to the choice of the filter width, and
it was decided to not apply any such filtering.  The frequencies in
common with both BRITE observing runs and their amplitudes are
included in Table~\ref{28freq}.  All values therein are averages over
the full range in time.

Figure \ref{28lcJDS} indicates the presence of some very slow regular
variability. In the frequency spectrum, there is a strong, well-isolated
peak at $f_f$ = 0.00471\,c/d.  The associated semi-amplitude is
16.7\,mmag, i.e., far higher than any of the variations seen with
BRITE.  A sine fit with this frequency is overplotted in
Fig.\,\ref{28lcJDS}.  Between phases 0.7 and 1.0, a similar bi-modal
distribution of values as in the annual light curve
(Fig.\,\ref{28lcSann}) is seen.  The variability may, therefore, be
due to an imperfect relative calibration of the cameras.  For
comparison, the SMEI data of some stars with similar positions in the
sky (\object{$\lambda$\,Cyg} [B5 V], \object{$\upsilon$\,Cyg} [B2
Vne], \object{QR\,Vul} [B3 Ve]) were analyzed.  Very strong
variability with 0.00411\,c/d was found in $\upsilon$\,Cyg and in QR
Vul with 0.00473\,c/d.  Accordingly, $f$ = 0.00471\,c/d is not thought
to be intrinisic to \object{28\,Cyg}.

The statistical properties of the SMEI data at their high-frequency
end are illustrated by Fig.\,\ref{deltaMagHistS} which presents a
histogram of the magnitude differences between immediately consecutive
orbits.  Comparison to the analogous Fig.\,\ref{deltaMagHistB} for BTr
shows that the stellar-to-instrumental ratio of the contributions to
these differences is smaller for SMEI (which has about the same
orbital period as the BRITE satellites have).  This difference is owed
to the systematics of the SMEI observations and the large noise from
the bright background due to zodiacal and scattered solar light.

\section{Analysis results}

\subsection{BRITE}
\label{BRITEresults}
\subsubsection{Common periodic variations in 2015 and 2016}
\label{BRITEcommon}

Figure \ref{28powerB} presents the frequency spectra of the BTr
observations in 2015 and 2016 (Fig.\,\ref{28LCf4}).  Beyond 3.5\,c/d,
no significant features can be found above the general noise floor.
Neither can obvious evidence for 1\,c/d and related peaks be seen.  A
global pictorial overview of the variability of 28\,Cyg as observed by
BRITE is provided by the two time-frequency diagrammes in
Fig.\,\ref{timefreq}.

\begin{table*}
  \caption{Mean seasonal values of frequencies and semi-amplitudes in
    \object{28\,Cyg}.  The list only contains frequencies detected with 
    BRITE in both 2015 and 2016 that agree to within 0.002\,c/d.} 
\label{28freq}
\centering
\begin{tabular}{l c c c c c c c c c}
\hline\hline       
Frequ.\ ID   & \multicolumn{2}{c}{BTr 2015} & \multicolumn{2}{c}{BTr 2016}  &  \multicolumn{2}{c}{BTr2014-2016}  & \multicolumn{2}{c}{SMEI} & {\sc Heros} \\
             & Frequ.    & Amplit.   & Frequ.    & Amplit.   & Frequ.    & Amplit.   & Frequ.    & Amplit.   & Frequ.     \\
             &   [c/d]   &   [mmag]  &   [c/d]   &   [mmag]  &   [c/d]   &   [mmag]  &   [c/d]   &   [mmag]  & [c/d]      \\
\hline
$f_{\Delta32}$, $f_{\Delta{cb}}$ &  0.0508   &    8.5    &   0.0496  &    12.8   &   0.0517  &    9.6    &  0.05093  &    2.3    &            \\
             &  1.3164   &    3.3    &   1.3177  &     5.4   &   1.3180  &    3.8    &           &           &            \\
             &  1.3603   &    2.2    &   1.3605  &     5.0   &   1.3599  &    3.0    &           &           &            \\
             &  1.3713   &    3.2    &   1.3691  &     4.2   &   1.3693  &    3.7    &  1.36809  &    2.3    &            \\
$f_1$, $f_a$ &  1.3799   &    6.8    &   1.3800  &     8.5   &   1.3811  &    6.6    &  1.37992  &    5.1    &             \\
$f_2$, $f_b$ &  1.5448   &    5.4    &   1.5446  &     5.5   &   1.5452  &    5.2    &  1.54470  &    3.3    & 1.54562     \\
$f_3$, $f_c$ &  1.5977   &    5.3    &   1.5976  &     6.5   &   1.5973  &    5.9    &  1.59695  &    1.3    & 1.59726     \\
             &  2.6769   &    3.9    &   2.6768  &     5.7   &   2.6761  &    4.5    &           &           &            \\
             &  2.7326   &    2.6    &   2.7319  &     4.4   &   2.7323  &    3.0    &           &           &            \\
             &  2.7407   &    2.5    &   2.7423  &     3.1   &   2.7420  &    2.5    & 2.74294   &     1.0   &            \\
$\approx f_2 + f_3$      &  3.1425   &    1.9    &   3.1427  &     1.9   &   3.1425  &    1.5    &           &           &            \\
\hline
\end{tabular}
\end{table*}

In the BRITE frequency spectra, the three highest peaks above 0.5\,c/d
occur at $f_1 = 1.381$\,c/d, $f_2 = 1.545$\,c/d, and
$f_3 = 1.597$\,c/d (here and in the following, all frequencies, etc.\
are from the combined analysis of the BTr observations in 2014, 2015,
and 2016 unless explicitly identified otherwise).  The strongest peak
below 0.5\,c/d appears at $f_{\Delta32} = 0.052$\,c/d.  This slower
variation is thought to be of initial relevance because the
preparation of the raw data did not include any correction for slow
variations.  Growing evidence for the significance of $f_{\Delta32}$
will emerge during the course of this subsection.  Despite of the much
shorter duration of the observations in 2014, $f_{\Delta32}$, $f_1$,
$f_2$, and $f_3$ were clearly detected also in this dataset alone.

Between the above frequencies, the following relation exists:

{\centering
0.0517\,c/d = $f_{\Delta32} \approx f_3 - f_2$ = 0.0521\,c/d

} 
\noindent
The deviation from full equality amounts to 0.0004\,c/d which is of
the order of the nominal single-season errors.  Figure \ref{28LCf4}
presents the light curve associated with $f_{\Delta32}$ in 2015 and
2016.



\begin{figure}
\includegraphics[width=6.2cm,angle=-90]{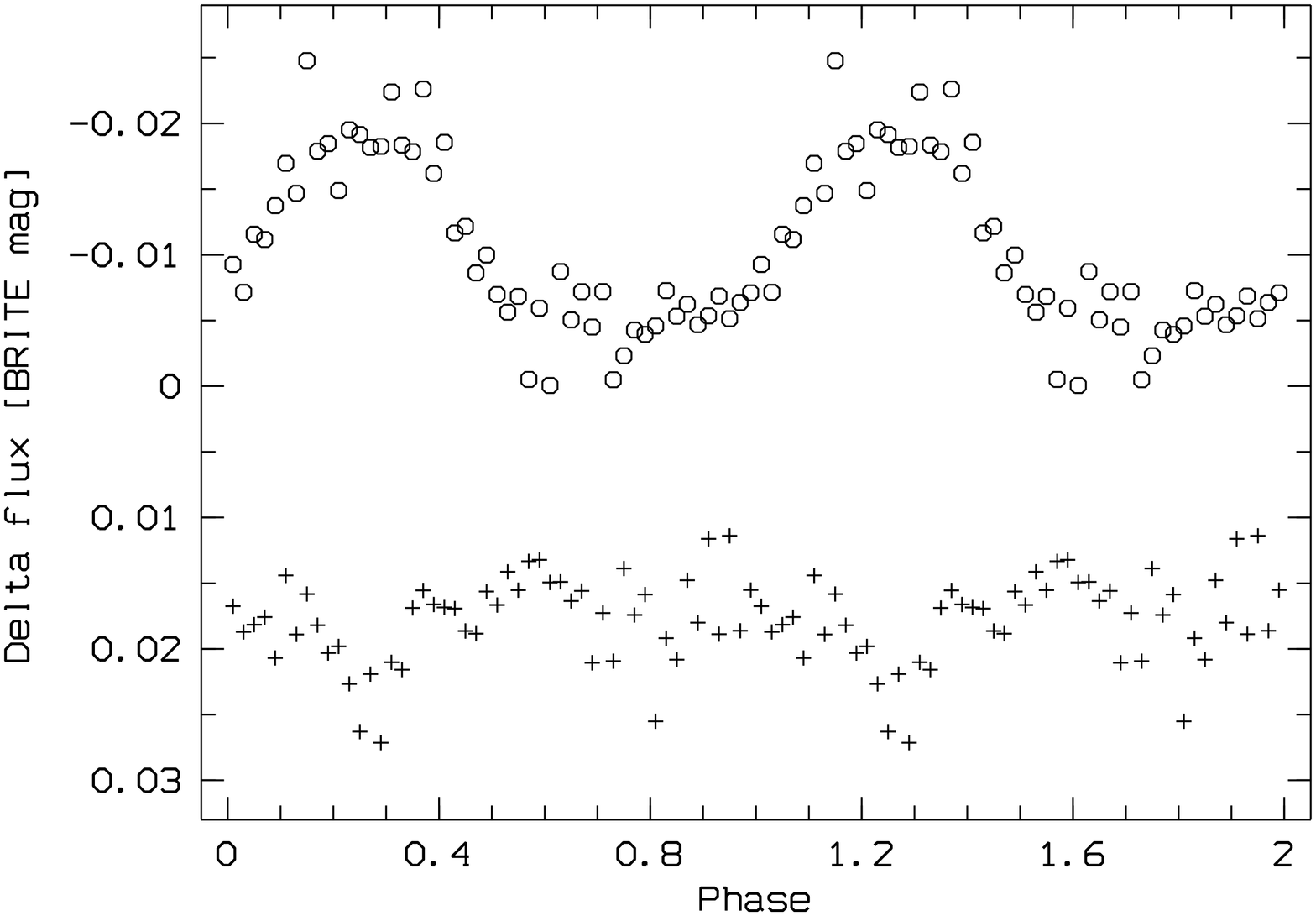}
\caption{
The BTr 2015 lightcurve from Fig.\,\ref{28LCf4} binned to steps of 
0.02 in phase of the 0.0508-c/d frequency ($f_{\Delta32}$).  Zero points 
of flux and phase are arbitrary.  For the same bins, the 
crosses indicate the scatter of the 2015 magnitudes in Fig.\,\ref{28LCf4}.  
Note the sign of the magnitude scale: The scatter is largest during 
light maxima.} 
\label{28LCf4scat2015} 
\end{figure}

\begin{figure}
\includegraphics[width=6.2cm,angle=-90]{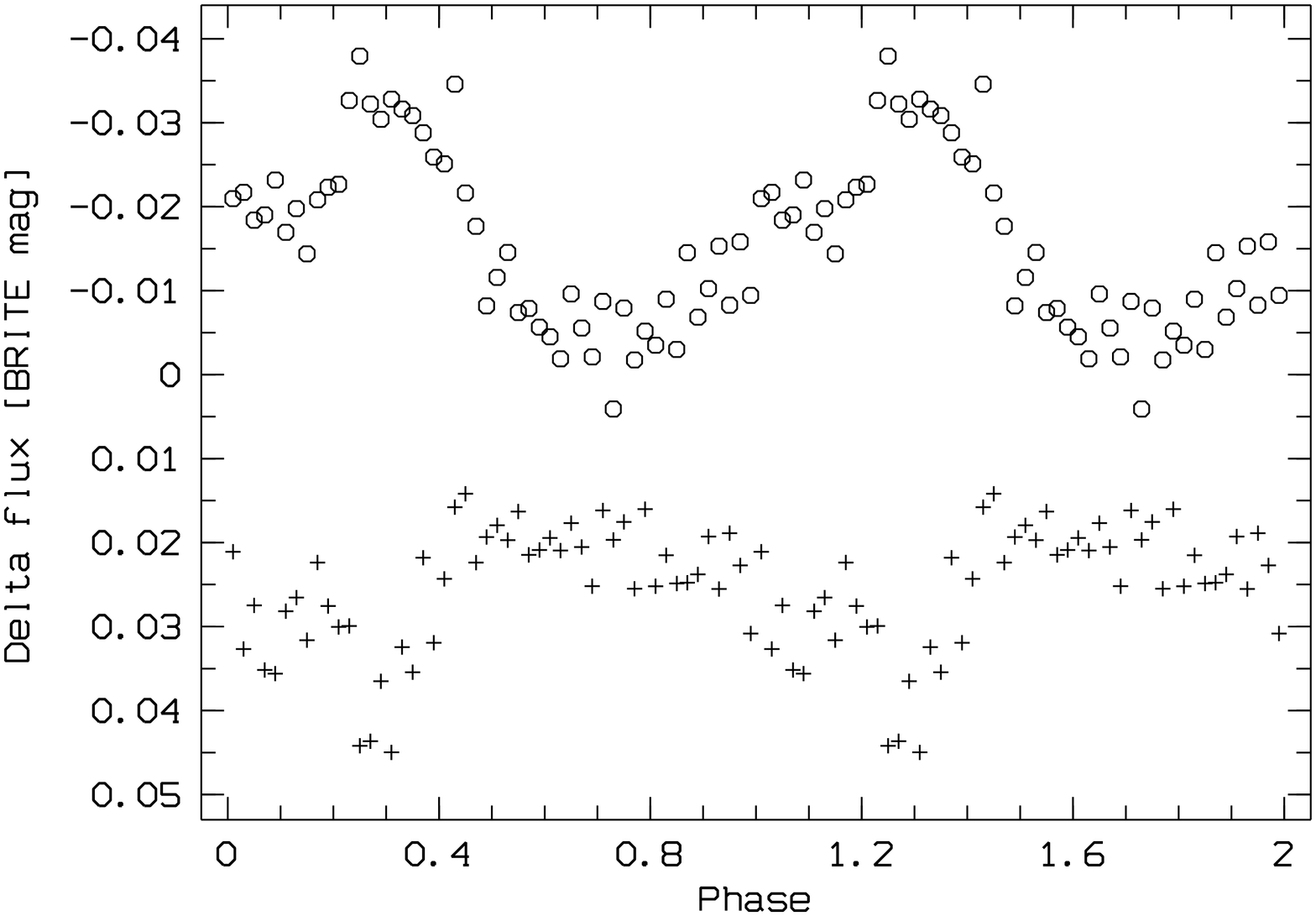}
\caption{
Same as Fig.\,\ref{28LCf4scat2015} except for BTr data from 2016 and 
$f_{\Delta32}$ = 0.0496\,c/d.  The offset in phase is arbitrary.  
} 
\label{28LCf4scat2016} 
\end{figure}

In 2015, 28\,Cyg did not undergo any obvious outburst signalling
mass-loss (Fig.\,\ref{28LCf4}).  However,
Fig.\,\ref{28LCf4scat2015} suggests small enhancements of the
photometric scatter during the maxima of the variability with
$f_{\Delta32} = 0.051$\,c/d.  The phase dependency of the scatter
became very visible (Fig.\,\ref{28LCf4scat2016}) in 2016.  The main
contribution to this was made by the last three $f_{\Delta32}$ cycles
in 2016, the amplitudes of which were up to an order of magnitude
higher than in the five preceding cycles (Fig.\,\ref{28LCf4}).  At the
time of these brightenings, the light curve also exhibited
interspersed short fadings well below the mean flux level.

Figure \ref{28powerB} shows also that the variability is clustered in three
frequency groups with approximate ranges 0.1-0.5\,c/d ($g_0$),
1.0-1.7\,c/d ($g_1$), and 2.2-3.0\,c/d ($g_2$).  The intervals between
them are not perfectly devoid of power spikes.  Contrary to $g_1$, the
frequencies in $g_2$ common to 2015 and 2016 do not belong to
the most prominent features of the group.  

The large number of temporary frequencies prompted a search for
frequency variations of $f_1$, $f_2$, $f_3$ and other frequencies in
single-season data.  For each frequency, a sine curve was fitted to
40-d intervals of the dataset concerned, and this time window was
shifted in steps of 1\,d.  No frequency appeared constant
(Fig.\,\ref{timefreq}).  However, lower-amplitude frequencies exhibited
higher scatter, and the shorter time interval of 40\,d implies larger
uncertainties in general.  Except for $f_1$ to $f_3$, which are
discussed in the following, the results are, therefore, mostly
inconclusive.

\begin{figure}
\includegraphics[width=6.6cm,angle=-90]{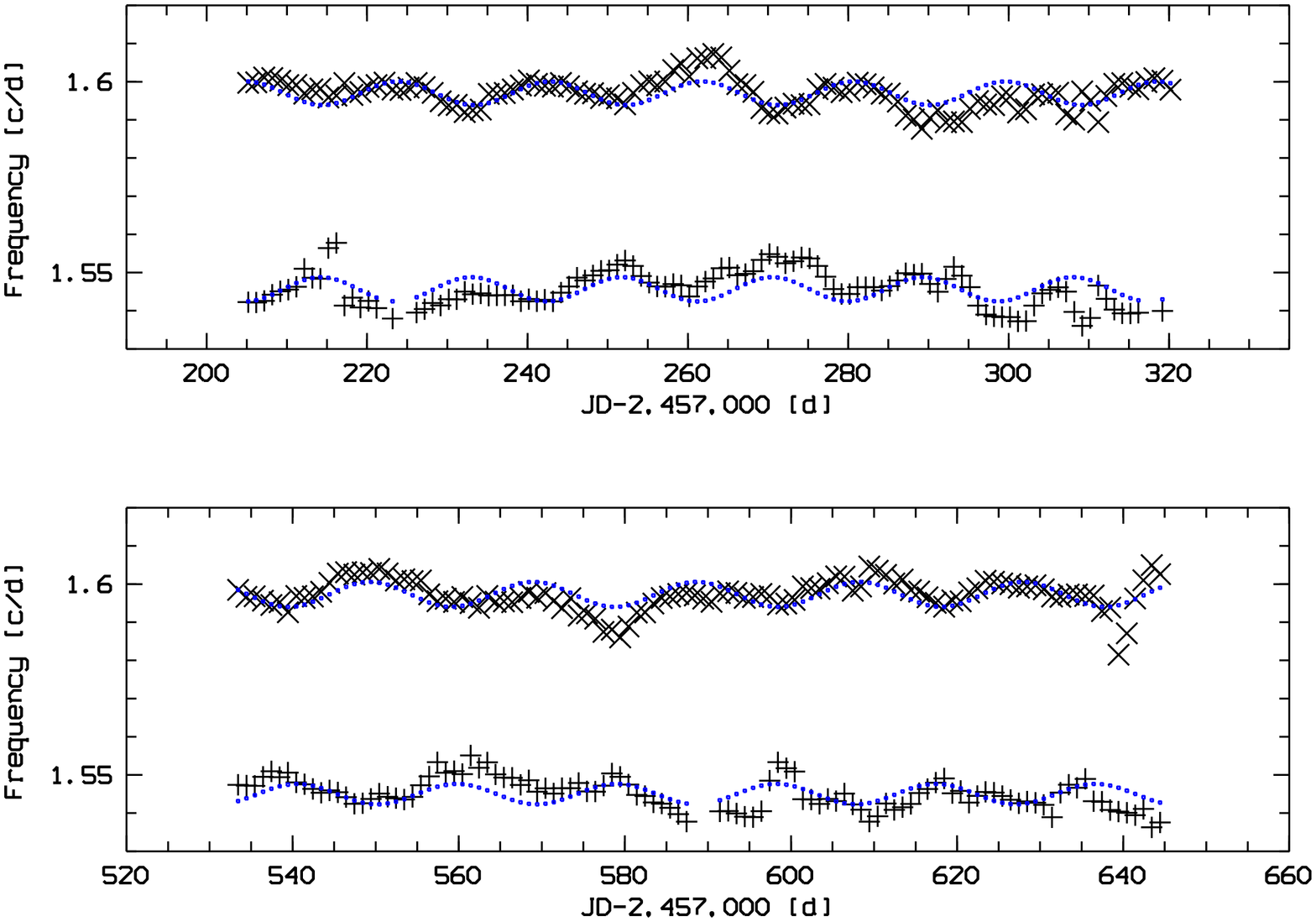}
\caption{ Variability of $f_2$ and $f_3$ in 2015 (top) and 2016
  (bottom).  Frequencies were measured in a window of 40\,d stepped by
  1\,d.  The overplotted blue dotted lines are fitted sine curves with
  frequency 0.0508\,c/d (2015) and 0.0496\,c/d (2016), tracing the
  respective mean $f_{\Delta32}$.  }
\label{VarFrq1516} 
\end{figure}

In both 2015 and 2016, $f_2$ and $f_3$ were modulated with their
$\Delta$ frequency, $f_{\Delta32}$.  The semi-amplitudes amounted to
0.005-0.007\,c/d, and $f_2$ and $f_3$ occurred nearly in antiphase
(Figs.\,\ref{timefreq} and \ref{VarFrq1516}).  There is no signature
of the large increase in amplitude of $f_{\Delta32}$ in the last two
months of the 2016 observations. To test the robustness of the result,
the time interval was varied between 20\,d and 60\,d in steps of
5\,d. There was no elementary change, and the antiphased modulation of
$f_2$ and $f_3$ was confirmed in all cases.  But at 60\,d the
amplitude was much reduced, and towards 20\,d $f_3$ and $f_2$ were no
longer properly resolved because, then, the length of the interval
corresponds to $1/(f_3 - f_2)$.

\begin{figure}
\includegraphics[width=6.6cm,angle=-90]{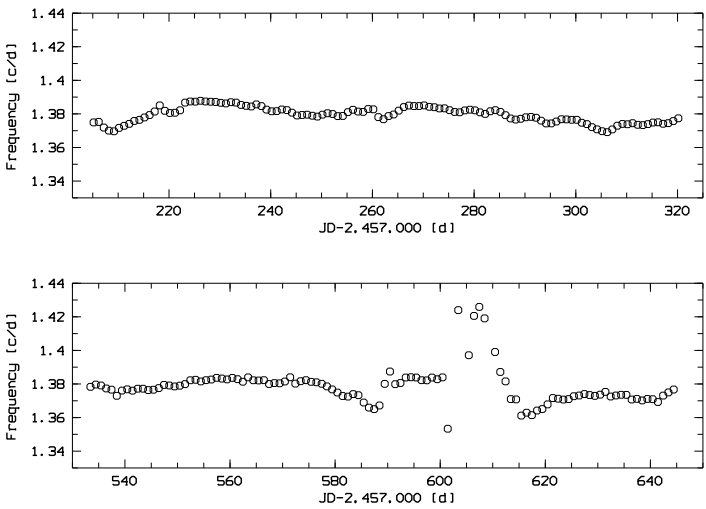}
\caption{
Variability of $f_1$ in 2015 (top) and 2016 (bottom). 
Frequencies were measured in a window of 40\,d stepped by 1\,d.  
} 
\label{VarFrq13800} 
\end{figure}

\begin{figure}
\includegraphics[width=6.6cm,angle=-90]{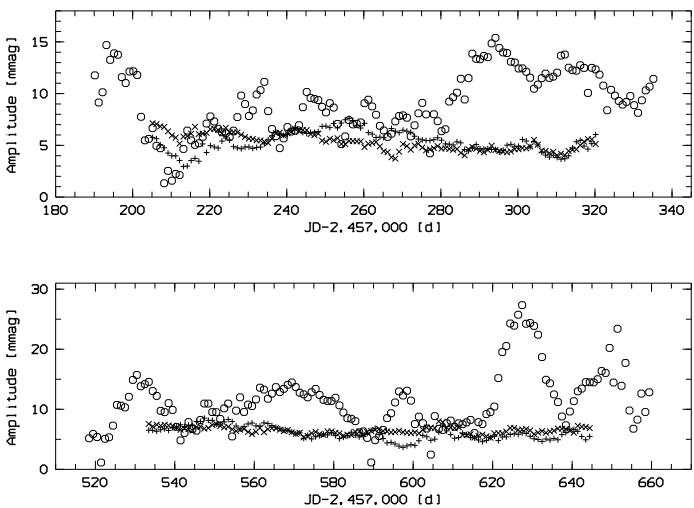}
\caption{
Variability in amplitude of $f_1$ ($\circ$), $f_2$ ($+$), and $f_3$ 
($\times$) in 2015 (top) and 2016 (bottom). 
Amplitudes were measured in windows of 10\,d ($f_1$) and 40\,d 
($f_2$ and $f_3$), stepped by 1\,d.  
} 
\label{VarAmp1516} 
\end{figure}

A similar frequency modulation does not affect the nearby $f_1$.
$f_1$ only suffered a major perturbation (Fig.\,\ref{VarFrq13800})
before the strong increase in amplitude of the $f_{\Delta32}$
variability around mJD = 610 (Fig.\,\ref{28LCf4}; mJD denotes a
modified Julian Date: HJD\,$-$\,2,457,000).  However, during that phase
of increased overall photometric amplitude, the amplitude of $f_1$
displayed semi-regular large-amplitude variations with a cycle length
comparable to, but clearly different from, that of $f_{\Delta32}$
(Fig.\,\ref{VarAmp1516}).  There was similar activity at the end of
the observations in 2015 (Fig.\,\ref{VarAmp1516}). By contrast, the
amplitude variations of $f_2$ and $f_3$ (Fig.\,\ref{VarAmp1516}) were
an order of magnitude lower than those of $f_1$.  The difference
between (i) the stellar variabilities with $f_2$ and $f_3$ and (ii)
the exophotospheric one with $f_1$ is very well visible in
Fig.\,\ref{timefreq}.

\begin{figure}
\includegraphics[width=6.7cm,angle=-90]{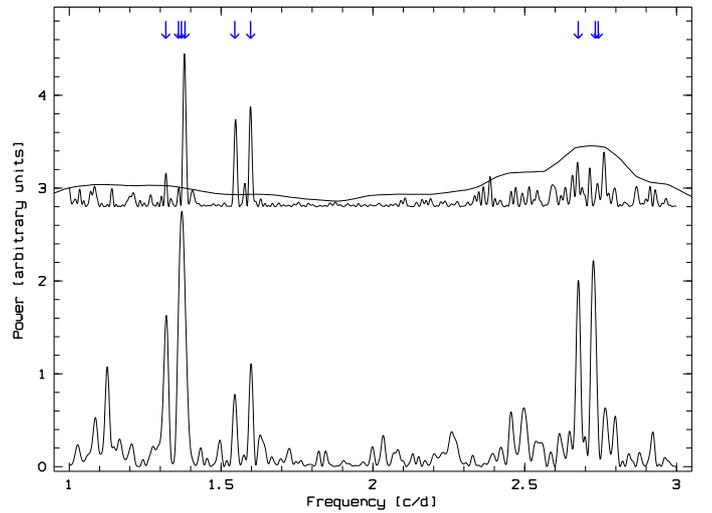}
\caption{ frequency spectra of the red 2016 BRITE observations before
  (top) and after (bottom) mJD\,600 when a series of brightenings
  began (see Fig.\,\ref{28LCf4}).  Power is given in arbitrary units.
  The arrows have the same meaning as in Fig.\,\ref{28powerB}, and the
  full-season error curve (arbitrarily scaled) from
  Fig.\,\ref{28powerB} is also shown.  In the upper panel, frequency
  groups $g_1$ (1.0-1.7\,c/d) and $g_2$ 2.2-3.0\,c/d are well
  separated.  }
\label{PS16lohi} 
\end{figure}

\subsubsection{Ephemeral variations}
\label{ephemeral}

Figure \ref{28powerB} gives rise to the suspicion that ephemeral
elements form a major part of the variability of \object{28\,Cyg}.
Motivated by the strong changes around mJD\,600, the 2016 data were
split into two sets, mJD$\leq$600 and mJD$>$600. The differences
between their frequency spectra are stunning (Fig.\,\ref{PS16lohi}).
The extra features during the outburst may be mixtures of (i)
combination frequencies, (ii) harmonics, (iii) side lobes due to the
large amplitude variation of $f_{\Delta32}$, (iv) circumstellar
variations and/or (v) stellar pulsations.  Since none of them was
detected in two or more independent datasets, their large number makes
it next to impossible to identify the nature of any such single
feature, and their short lifespans prevent their characterization as
periodic.

Temporarily enhanced visibility is not necessarily the same as
increased amplitude or power.  In very rapidly rotating stars, the
angular parts of NRP eigenfunctions are described by a series of
spherical harmonics \citep{2008ApJ...679.1499L}.  If the decomposition
of eigenmodes into spherical harmonics changes during an
outburst/brightening, this could also modify their visibility.  In
that case, such modes would not add to the cause of
outbursts/brightenings.  At the current level of analysis, only
$f_{\Delta32}$ can be considered as causing mass loss.

The variability of Be stars is not erratic or even chaotic: The clear
signature of low-order $g$-mode pulsations in high-resolution spectra
\citep{2003A&A...411..229R} demonstrates coherent, and coherently
varying, large-scale structures in the photosphere.  Moreover, for
several dozen spikes in the frequency spectrum the data were convolved
with them, and in most cases the resulting light curve had an
approximately sinusoidal shape.

Finally, the frequency spectra of the 2015/16 BTr data were searched
for patterns by calculating histograms of nearest-neighbour
differences, frequency spectra (of the frequency spectra), and
cross-correlation functions.  The results were all negative, except
for a possible overabundance of differences close to 0.01\,c/d (e.g.,
1.369\,c/d - 1.360\,c/d and 2.742\,c/d - 2.732\,c/d in
Table\,\ref{28freq}; 1.369 and 2.742\,c/d are harmonics).  In the SMEI
frequency spectrum, a feature exists at 0.00995\,c/d; at 8.6\,mmag,
the amplitude is large.  A similar light curve arises from the BTr
observations in 2016.

\subsection{SMEI}
\label{SMEIanalysis}

\begin{figure}
\includegraphics[width=6.7cm,angle=-90]{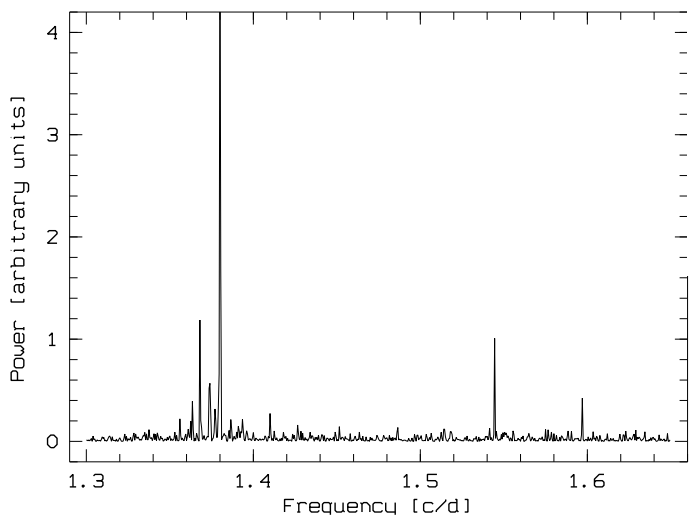}
\caption{
frequency spectrum (in arbitrary units) of the SMEI data in the 
region of the main BRITE frequencies above 0.5\,c/d ($f_a$: 1.381\,c/d, 
$f_b$: 1.545\,c/d, and $f_c$: 1.597\,c/d). 
} 
\label{28powerNarrowS} 
\end{figure}

\begin{figure}
\includegraphics[width=6.5cm,angle=-90]{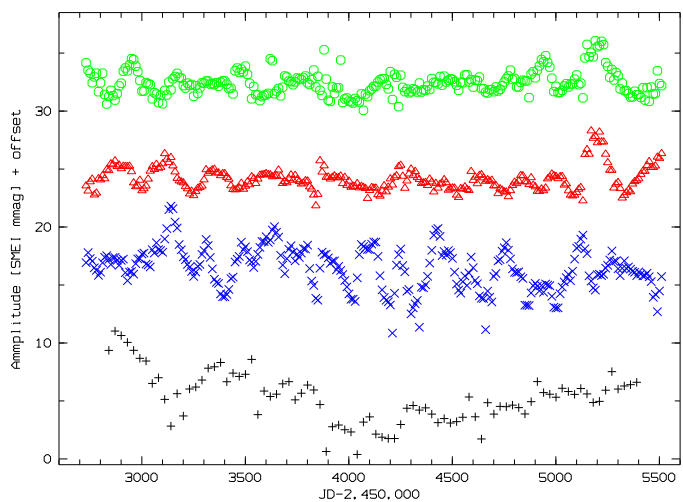}
\caption{ Time-amplitude diagram for $f_{\Delta32}$ (black $+$), $f_1$
  ($\times$), $f_2$ (triangles), and $f_3$ ($\circ$) in the SMEI data.
  The semiamplitudes were derived from fits of single sine functions
  over windows of 100\,d ($f_{\Delta32}$: 300\,d) stepped by 10\,d
  ($f_{\Delta32}$: 30\,d).  From bottom to top, the curves are
  vertically shifted by 0, 10, 20, and 30\,mmag, respectively.  }
\label{VarAmpS} 
\end{figure}

\begin{figure}
\includegraphics[width=6.5cm,angle=-90]{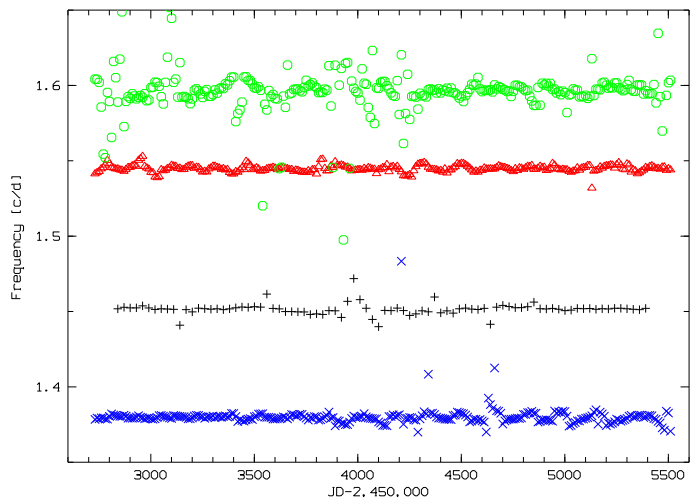}
\caption{
Same as Fig.\,\ref{VarAmpS} except for time-frequency diagram.  
The frequency curve of $f_{\Delta32}$ ($\sim$0.05\,c/d) was vertically 
shifted by 1.4\,c/d.  
} 
\label{VarFrqS} 
\end{figure}

The three main BRITE frequencies above 0.5\,c/d stand out also in the
SMEI data (Fig.\,\ref{28powerNarrowS}).  Peaks at $f_a = 1.3799$\,c/d,
$f_b = 1.5447$\,c/d, and $f_c = 1.5970$\,c/d can be identified with
$f_1$, $f_2$, and $f_3$ (Table \ref{28freq}).  That is, the mean
frequencies in 2003-2010 (SMEI) as well as in 2014-2016 (BRITE) were
the same.  The long duration of the SMEI observations permits the
constancy of the frequencies to be examined on shorter timescales.
The results of single sine curves fitted to the data
(Fig.\,\ref{VarAmpS}: amplitudes; Fig.\,\ref{VarFrqS}: frequencies)
exhibit some commonalities of the amplitude variations of $f_2$ and
$f_3$ whereas the behaviour of $f_1$ is different and more pronounced.
With a factor of $\sim$3, the largest relative amplitude variation was
associated with $f_{\Delta32}$.

In the SMEI observations, a similar relation exists between 
$f_a$, $f_b$, and $f_c$ as between $f_1$, $f_2$, and $f_3$ with BRITE:

{\centering

0.05093\,c/d = $f_{\Delta{cb}} \approx f_c - f_b$ = 0.05225\,c/d

}

The light curve after prewhitening for 0.00471\,c/d
(Sect.\,\ref{SMEIobs}) suggests that SMEI did not capture outbursts
larger than 0.05\,mag and longer than 100\,d.  This is in agreement
with, but less stringent than, the BRITE observations
(Sect.\,\ref{BRITEresults}) and the findings of
\citet{2001PASP..113..748P}. Even after this correction, no hint of a
photometric counterpart of the fading of the H$\alpha$ emission from
2005/2007 to 2013 (Sect.\,\ref{spectra}) is present.

\subsection{Spectroscopy} 
\label{spectra}

\citet{2000ASPC..214..232T} (see also Sect.\,\ref{28cygintro}) found
two frequencies, $\nu_1 = 1.54562$\,c/d and $\nu_2 = 1.60033$\,c/d
with radial-velocity amplitudes of $\sim$20 and $\sim$10\,km/s,
respectively.  If corrected for a likely 1-c/yr alias, $\nu_2$ becomes
1.59726\,c/d, and $\nu_1$ and $\nu_2$ can be perfectly identified with
BRITE frequencies $f_2$ = 1.5453\,c/d and $f_3$ = 1.5973\,c/d,
respectively.  The associated line-profile variability seems to be of
the same $\ell = -m = 2$ $g$-mode variety that is typical of classical
Be stars \citep{2003A&A...411..229R}.  Accordingly, $f_2$ and $f_3$
are due to nonradial pulsation $g$ modes.

\begin{figure}
\includegraphics[width=3.4cm,angle=-90]{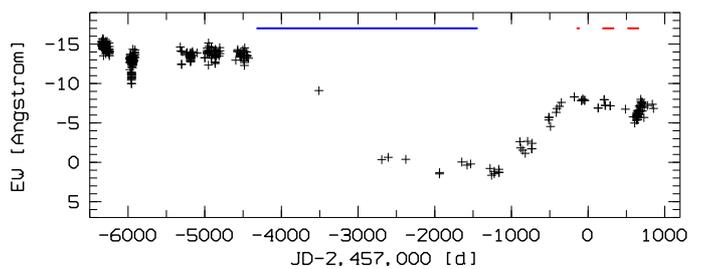}
\caption{H$\alpha$ equivalent widths in {\sc Heros} \citep[][and
  unpublished]{2000ASPC..214..232T} (JD < 2,453,000) and BeSS
  \citep{2011AJ....142..149N} (JD > 2,453,000) spectra.  Negative
  values mean net line emission.  The 1-$\sigma$ scatter around
  JD\,2,452,000 is 0.5\,\AA; for BeSS errors see Fig.\,\ref{VtoR}.
  The short red horizontal bars near the upper right corner mark the
  BRITE observations.  The long blue bar spans the SMEI
  observations.}
\label{28CygHalphaBeSS} 
\end{figure}

As of February 2017, the BeSS database \citep{2011AJ....142..149N}
contained 118 wavelength-calibrated H$\alpha$ line profiles
contributed by more than a dozen amateurs between 1995 July and 2017
January, about half of them in 2016. Those observed since August 1997
were downloaded from the BeSS Web site.  The equivalent widths derived
from them are presented in Fig.\,\ref{28CygHalphaBeSS}.  Like the
photometry, the contemporaneous H$\alpha$ spectroscopy conveys the
picture of 28\,Cyg as one of the more quiet early-type Be stars.  The
H$\alpha$ line emission was at a crudely constant high level through
the end of 2002.  From mid-2007 until early 2012 it was weak but again
fairly constant.  Thereafter it started to rise but did not reach the
previous high level.  The BRITE photometry started nearly 3 years
after the onset of this partial recovery.  During the BRITE
observations the line emission stayed at some plateau with little
variability.  The overall timescale is at the long end of such cycles
\citep[e.g.,][]{2016ASPC..506..315G}.

\begin{figure}
\includegraphics[width=6.2cm,angle=-90]{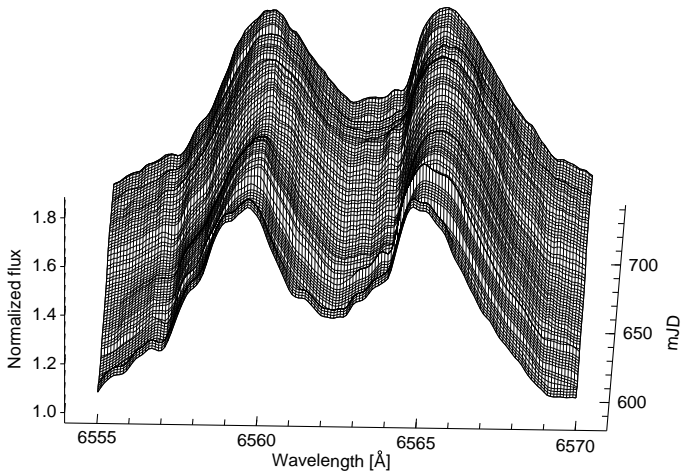}
\caption{
Combination of 60 BeSS H$\alpha$ profiles between mJD\,589 and 733. 
The spectra were averaged over a window of 20\,d stepped by 2\,d.  The 
spectral bin size is 0.1\,\AA.  Only three spectra were obtained before 
mJD\,633.  The BRITE observations terminated on mJD\,665.  
} 
\label{BeSSperspHalpha} 
\end{figure}

\begin{figure}
\includegraphics[width=10.5cm,angle=-90]{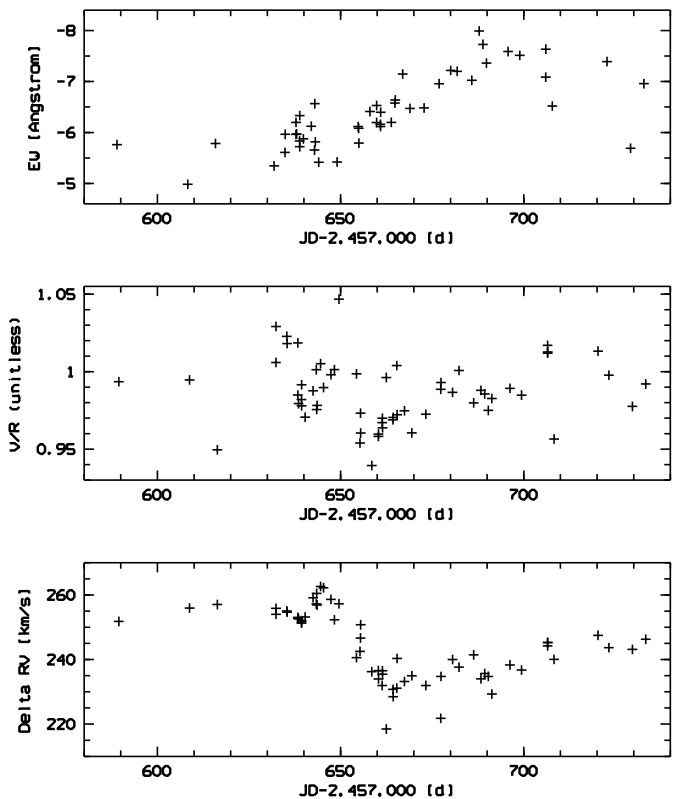}
\caption{ Variability in total equivalent width (top,
  cf.\,Fig.\ref{28CygHalphaBeSS}) peak-height ratio (middle), and peak
  separation (bottom) of the violet and red H$\alpha$ emission peaks
  in Fig.\,\ref{BeSSperspHalpha}.  Estimated 1-$\sigma$ errors are
  0.4\,{\AA}, 0.02, and 4\,km/s, respectively.}
\label{VtoR} 
\end{figure}

There are no BeSS spectra from 1998 when \citet{2000ASPC..214..232T}
observed an outburst with {\sc Heros}.  A single H$\alpha$ profile
20\,d before the amplitude maxima of $f_a$, $f_b$, and $f_c$ seen by
SMEI (Sect.\,\ref{SMEIanalysis}) is free of anomalies.  Of the 62 BeSS
spectra from 2016, 57 were obtained between mJD\,630 and 740 in
response to an alert when the online control of the BRITE observations
had recognized the increased activity starting around mJD\,600.  The
spectra were manually normalized with fitted splines, resampled from
their original wavelength bins between 0.03\,{\AA} and 0.18\,{\AA} to
a common 0.1\,\AA, and are plotted in Fig.\,\ref{BeSSperspHalpha}.

The two H$\alpha$ emission peaks are well, albeit not perfectly,
approximated by Gaussians so that differential measurements from
fitted Gaussians make sense.  Figure\,\ref{VtoR} presents the
variability in total equivalent width, peak-height ($V/R$) ratio, and
peak separation measured in the BeSS spectra; because of the
inhomogeneous provenience of the observations only rough error
estimates can be given.  After mJD\,650, there was a gentle increase
by 25\% in equivalent width.  At about the same time, a fairly marked
drop by more than 10\% occurred in the peak separation.  No attempt
was made to search for repetitive variability on any timescale.

From 44 ground-based spectra \citet{1988ESASP.281b.117P} derived an
NRP period of 16.5\,h (f: 1.45\,c/d), which is a blend of $f_1$-$f_3$,
and found it co-phased with the equivalent width of
C\,IV\,$\lambda\lambda$\,1548,1551 in 17 $R \approx 10,000$ spectra
from the {\it International Ultraviolet Explorer} ({\it IUE})
\citep{1978Natur.275..372B}.  In 25 {\it IUE} spectra obtained within
56\,h, \citet{2000ASPC..214..375P} found a period of
0.646$\pm$0.019\,d (1.55$\pm$0.05\,c/d, i.e., very close to $f_2$ =
1.545\,c/d and still consistent with $f_3$ = 1.600\,c/d).  Because the
flux amplitude increased towards lower wavelengths, it was attributed
to pulsational temperature variations.  The C\,IV\,$\lambda$\,1551
wind line exhibited correlated variability in equivalent width while
the photospheric C\,III\,$\lambda$\,124.7\,nm and
Si\,II\,$\lambda$\,126.5\,nm lines did not.  However, Peters \& Gies
did not state whether the variability in line strength was above the
escape velocity so that the relation to genuine mass loss is not
clear.

\subsection{Caveats and numerical limits}
\label{caveats}

The analysis of the photometry of 28\,Cyg is hitting a fundamental
mathematical limit: To measure a frequency $f$, a conservative rule of
thumb is that the data should cover a time interval T\,$\geq$\,2/$f$.
In the presence of noise, intervals larger by an order of magnitude
are desirable.  If the case of closely spaced power peaks, which,
moreover, undergo independent variations, this minimal duration will
often not suffice.  The timescales of frequency and amplitude
variations in 28\,Cyg are comparable to this safe width of time
windows.  This touches upon the definition of a frequency.

Figure\,\ref{timefreq} illustrates the complexity of the
variabilities, their variations in both frequency and power, in a
single picture.  For the numerous secondary frequencies
(Sect.\,\ref{ephemeral}), error and significance analyses attain a
very limited meaning in the presence of the numerous interrelations.

The remainder of the paper only discusses frequencies $f_1$, $f_2$,
$f_3$, and $f_{\Delta32}$ (Table\,\ref{28freq}) all four of which were
detected by BRITE in 2015 as well as in 2016
(Sect.\,\ref{BRITEresults}) and confirmed with SMEI
(Sect.\,\ref{SMEIanalysis}).  Moreover, $f_2$ and $f_3$ were first
spectroscopically discovered.  The maximal differences between the
photometric values of these four frequencies are 0.0001, 0.0002,
0.0007, and 0.0013\,c/d, respectively.  Figure\,\ref{timefreq}
suggests a frequency jitter above that level.  Therefore,
Table\,\ref{28freq} only contains frequencies detected in both BRITE
datasets (from 2015 and 2016) and agreeing to better than
$\sim$0.002\,c/d. As will become amply clear below, only very few of
the many processes in 28\,Cyg, if any, seem to be running on clocks
with that precision.  Moreover, the entries in Table\,\ref{28freq} 
are seasonal averages.  There are only eleven such seemingly persistent
frequencies; they are marked in Fig.\,\ref{28powerB}.

A low-threshold search between 0 and 3.5\,c/d at a sampling of
0.001\,c/d found $\geq$60 frequency peaks in 2015 and $\geq$100 in
2016.  For randomly distributed frequencies, the probabability of
chance coincidences within 0.002\,c/d is, then, less than 6\%.  It is
lower for the much fewer high-amplitude variabilities of
Table\,\ref{28freq}.  For a set of n lines in two independent
observations (BRITE and SMEI), the single-frequency probability is
raised to the nth power.  The dashed lines in Fig.\,\ref{28powerB}
mark 3-$\sigma$ noise levels after subtraction of these variabilities.
This rigorous approach does not take into account the large amplitude
variability of $f_{\Delta32}$ especially at the end of the
observations from 2016 (Fig.\,\ref{28LCf4}).  Moreover, it attributes
all other variability completely to stellar or instrumental noise.
Such a classification system may be too crude for the variability of
\object{28\,Cyg} (Sect.\,\ref{ephemeral}).

For reference, we mention the work of \citet{2017A&A...603A..13K} who
determined a total of 22 frequencies with errors between 0.9 and
92~$10^{-5}$\,c/d and amplitudes from 0.6 to 17\,mmag from BRITE
observations of the SPB star HD\,201433.  SMEI and BRITE frequencies
agreed at the 0.8 and 1.8\,$\sigma$ level, respectively.  In the
longer SMEI observing sequences, 9 rotationally split frequency pairs
could be resolved.

Given the large PSFs of SMEI and BRITE, there is the risk that the
main frequencies $f_1$, $f_2$, $f_3$, and $_{\Delta32}$ are not
intrinsic to \object{28\,Cyg} but are contributed by one or more other
stars.  Any such star would need to have fallen into the apertures of
both SMEI and BRITE.  The radius of a circle including the PSF at both
nodding positions is $\sim$12\,arcmin.  As of August 2017, the
International Variable Star Index \citep{2006SASS...25...47W} in
VizieR \citep{2000A&AS..143...23O} contains 2 variable stars within
15\,arcmin from \object{28\,Cyg}.  One (VSX J200906.4+365552) is an
M5\,V star that is listed as varying in $R$ between 8.4 and 8.8\,mag
with unknown period.  For the other one (VSX J201000.6+365111), too, a
period is not known, and the stated $R$ magnitude range is
11.8-12.4\,mag.

In principle, this still leaves dozens of other objects as
hypothetical real sources of the observed frequencies.  However, since
$f_{\Delta32}$ is the difference between $f_3$ and $f_2$, these three
frequencies must be from one and the same star.  Moreover, $f_2$ and
$f_3$ were detected spectroscopically so that the common source of
these three frequencies was in the aperture of the spectrograph.  Only
$f_1$ could be from a second star; however, there are plausible
reasons (Sect.\,\ref{periodic}) to identify it as the {\v S}tefl
frequency of this Be star.

\section{Discussion}
\label{discussion}
\subsection{Conceptual framework}
\label{ansatz}

Most Be stars undergo large-scale variations of their discs on
timescales of years.  Changes in total disc mass manifest themselves
in varying H$\alpha$ emission-line strengths and photometric
amplitudes of several 0.1 mag \citep[e.g.,][]{2002AJ....124.2039K}.
Evidence of more nearly periodic slow photometric variability in Be
stars is still restricted to shorter timescales.
\citet{2016arXiv160908449L} present light curves of 11 Be stars with
apparent periods >80\,d.  However, it is impossible to recognize any
Be star by specific properties of this variability.  Perhaps, this
informational shallowness is the reason why the key to the
understanding of the Be phenomenon has not been found therein.

Superimposed on this slow variability are outbursts.  Their
spectroscopic record is particularly rich and multifacetted for
\object{$\mu$\,Cen}, reaching from pulsationally triggered mass
ejections \citep{1998ASPC..135..343R} via new matter settling in the
circumstellar disc \citep{1988A&A...198..211B, 1991ESOC...36..185H} to
gas moving towards the star at high velocity
\citep{1986ApJ...301L..61P}.  \citet{2002A&A...388..899N} also
captured an outburst in spectra of \object{$\omega$\,Ori}.

The build-up of line emission, i.e., a disc, as well as
super-equatorial velocities associated with {\v S}tefl frequencies
\citep{1999MNRAS.305..505S} are evidence of star-disc mass transfer.
There may also be numerous failed mass-loss events, during which
matter is tossed up but does not reach an orbit as suggested by rapid
variability in broad-band polarimetry \citep{2007ApJ...671L..49C}.
Such cases are not easily diagnosed by spectroscopy but may still emit
additional light or remove light from the line of sight, or both.
Therefore, the following will speak of `brightenings' and leave it
open from which level on they may be photometric equivalents of
spectroscopic mass-loss events.  Inspection of the light curve
(Fig.\,\ref{28LCf4}) shows that 28\,Cyg quite regularly experiences
brightenings with amplitudes well above the higher-frequency
variability floor.

For the interpretation of apparent exophotospheric photometric
variability, the model calculations by \citet{2012ApJ...756..156H} are
used as the foundation.  They imply that at an inclination angle near
$70^{\circ}$ the photometric variability due to emission from variable
amounts of near-stellar matter vanishes.  On this basis, the estimates
of the inclination angle of 28\,Cyg, $40^\circ < i < 75^\circ$
(Sect.\,\ref{28cygintro}), suggest that matter ejected into the
equatorial plane produces extra emission that ought to reveal itself
as brightenings (see also Fig.\,\ref{VDD}, top panel). However,
the sensitivity may be small.  In the wavelength region covered by
BRITE Constellation, the matter involved is expected within about two
radii of (one radius above) the stellar photosphere
\citep{2012ApJ...756..156H}.  Photometry alone cannot determine
whether this matter is already part of the disc, drifting towards it,
or falling back to the star.  It may form one large contiguous
structure or be fractionated into numerous small blobs.  Therefore,
the following will use the term `exophotospheric emission regions'
(EERs) to allude to such situations.

The aspect-angle dependence may be different if matter were also lost
at higher stellar latitudes \citep[as also considered
by][]{2003A&A...402..253S, 2010ApJ...709.1306W}.  The ejecta may
initially subtend an angle comparable to or smaller than the
photosphere so that they absorb photospheric radiation as they quickly
cool radiatively below the photospheric temperature.  With time, the
material spreads out, becoming much less dense, and partly reaches a
Keplerian orbit.  Most of this matter will not be seen projected on
the stellar disc but, while close to the star, produce a brightening
resulting from the reprocessing of stellar radiation.

To date, in the BRITE sample of Be stars, only \object{$\eta$\,Cen}
has been examined \citepalias{2016A&A...588A..56B} in similar detail
as \object{28\,Cyg}.  In \object{$\mu$\,Cen}
\citepalias{2016A&A...588A..56B} strong emission from irregularly
variable EERs seen at high inclination prevented detection of the
stellar variability.  The top-level variability of
\object{$\eta$\,Cen} comprises the following: (i) Two
spectroscopically confirmed $g$ modes, (ii) their $\Delta$ frequency
with a much larger amplitude than the sum of the two $g$-mode
amplitudes, and (iii) a so-called {\v S}tefl frequency which is
exophotospheric in nature (cf.\ Sect.\,\ref{circumstellar}),
$\sim$10\% slower than the $g$-modes, and has a very large amplitude.
\citetalias{2016A&A...588A..56B} has argued that the $\Delta$
frequency modulates the star-to-disc mass transfer and the amplitude
of the {\v S}tefl frequency traces the degree of orbital
circularisation of matter in the inner disc.  A transient 0.5-c/d
variability that \citet{2002A&A...388..899N} attribute to a cloud
revolving \object{$\omega$\,CMa} for a few orbits may be of a similar
nature.

The structural similarity of the variabilities of $\eta$\,Cen and
28\,Cyg is easily recognizable.  28\,Cyg, too, features two $g$ modes
($f_2$ and $f_3$), their $\Delta$ frequency $f_{\Delta32}$, whose
average amplitude is not much smaller than those of $f_2$ and $f_3$
together, and $f_1$, which is $\sim$10\% lower than $f_2$ and $f_3$
but has the largest mean amplitude (see Fig.\,\ref{VarAmpS}).  The
structure of this section follows the structure of the variability.
The stellar variability is evaluated in Sect.\,\ref{periodic}.
Section \ref{massloss} compiles the evidence that $f_{\Delta32}$
regulates the mass loss, and Sect.\,\ref{circumstellar} examines $f_1$
as a circumstellar {\v S}tefl frequency.  The discussion begins with
the slow irregular variability.

\subsection{Slow and irregular variability}
\label{nonperiodic}

The clearest indicator of structural variations of Be discs is the
violet-to-red ratio of emission peaks, $V/R$.  It is a robust
estimator as it only requires a differential measurement.  In
high-inclination stars like 28 Cyg, the large rotational peak
separation makes $V/R$ variations easily detectable.  It can be
induced on very different timescales as global disc oscillations by
the quadrupole moment of a rotationally distorted central star
\citep{1997A&A...318..548O}, dynamically by any sufficiently massive
and close companion \citep{2007ASPC..361..274S, 2017arXiv170406751P},
radiatively by a hot companion \citep[e.g., ][]{2013ApJ...765....2P,
  2015A&A...577A..51M}, and by non-axisymmetric mass injections.  Also
in this respect, 28\,Cyg is at the far quiet end of the activity scale
of early-type Be stars.

The persistent high degree of symmetry of the H$\alpha$ line emission
suggests that any undetected companion star is too ineffective in
affecting the disc.  Therefore, the explanation of the slow
equivalent-width variations of the H$\alpha$ profiles
(Fig.\,\ref{28CygHalphaBeSS}), which leave no doubt that the disc
around 28\,Cyg was eroded for some years and was replenished around
2013, must be sought in a single-star context.  Contrary to many years
of ground-based photometric monitoring, \citet{2000ASPC..214..232T}
recorded a discrete mass-loss event in 1998, when the absolute value
of the H$\alpha$ equivalent width dropped by $\sim$30\%.  The event
lasted 10-15\,d and was the only one observed in two seasons.  The
low level of disc activity is confirmed by the 5-months BeSS snapshot
(Figs.\,\ref{BeSSperspHalpha} and \ref{VtoR}), during which no such
event was observed (i.e., after the brightening starting around
mJD\,600).  The slow variability is probably due to oscillatory disc
instabilities \citep{1997A&A...318..548O}.

The low variability in line emission is noteworthy because in the VDD
model, without replenishment, Be discs are re-accreted within months
to years \citep{2012ApJ...744L..15C, 2017MNRAS.464.3071V}.  Radiative
ablation may accelerate this process \citep{2016MNRAS.458.2323K}.
Therefore, one might expect the H$\alpha$ line emission to be fairly
variable.  The simplest resolution of this puzzle may consist of
quasi-continuous star-to-disc mass transfer, which could also include
small outbursts every couple of weeks.

\subsection{The $g$ modes $f_2$ and $f_3$ and other stellar 
variabilities}
\label{periodic}

The most remarkable fact about the variability of 28\,Cyg is the tight
link, via $f_{\Delta32}$, between the two $g$ modes with $f_2$ and
$f_3$.  Neither the reason nor the geometric depth of the combination
is known.  It could be just a linear superposition and merely appear
as a coupling due to the extra emission from the mass ejections it
causes, perhaps amplified by a nonlinear response of the atmosphere to
this slow and possibly non-adiabatic variability.  There could also be
non-linear mode coupling deep inside the star
\citep{1982AcA....32..147D}.

Two closely spaced frequencies also form part of an anchor variability
pattern (cf.\ Sect.\,\ref{ansatz}) in many other (early-type) Be
stars.  Examples include \object{$\eta$\,Cen}
\citepalias{2016A&A...588A..56B}, \object{$\psi$\,Per},
\object{25\,$\psi^1$\,Ori} \citep{2017ASPC..508...93B},
\object{10\,CMa}, and \object{27 CMa}
\citep{2016arXiv161101113B}. However, in pole-on stars, the light
curves are dominated by emission from varying amounts of near-stellar
matter that blanket any underlying stellar variability.
\object{$\mu$\,Cen} \citepalias{2016A&A...588A..56B},
\object{$\kappa$\,CMa}, and \object{$\omega$\,CMa}
\citep{2016arXiv161101113B} are such cases.  Therefore, the fraction
of (early-type) Be stars with observed photometric $\Delta$
frequencies will always be lower than the genuine incidence of
$\Delta$ frequencies.  If many Be stars exhibit large-amplitude
$\Delta$ frequencies, the conditions for their formation should not be
too special.

$f_2$ and $f_3$ are the most stable among the frequencies investigated
(Figs.\,\ref{timefreq}, \ref{VarFrq1516}, and \ref{VarFrqS};
Table\,\ref{28freq}).  However, they undergo a subtle modulation with
$f_{\Delta32}$, i.e., their own difference.  Not any less mysterious
is that $f_2$ and $f_3$ vary in antiphase (Figs.\,\ref{timefreq} and
\ref{VarFrq1516}).  This may harbour some hint as to how and where the
coupling of these two NRP modes takes place.  While it is still
conceivable that the $\Delta$ frequency modulates the position of the
layer where the waves associated with the $g$ modes are reflected and
so modulates their apparent frequencies, it is not at all clear why
two very closely spaced frequencies are affected in antiphase.

Also the constancy in amplitude of $f_2$ and $f_3$ stands out above
much of the rest (Figs.\,\ref{timefreq}, \ref{VarAmp1516}, and
\ref{VarAmpS}).  If the mass-loss modulation and apparent driving with
$f_{\Delta32}$ (Sect.\,\ref{massloss}) were ultimately powered by
these two $g$ modes, one might not expect this.  Comparison to
groundbased studies shows that most of them found some blend of $f_1$,
$f_2$, and $f_3$ but could not resolve it.  The implied apparent 
variability of the frequencies contributed to the
notion prevailing at that time that the rapid variability of Be stars
is mostly semi-regular.  Thanks to BRITE and SMEI it is now clear 
that stellar variabilities can persist for decades 
\citepalias[see also][]{2016A&A...588A..56B}.

\subsection{$f_{\Delta32}$: The brightening in 2016 and other
  mass-loss indicators}
\label{massloss}
$f_{\Delta32}$ was not detected by ground-based observations;
0.09\,c/d reported by \citet{1994IAUS..162..102R1997A&AS..125...75P}
is close to the first harmonic.  $f_{\Delta32}$ is the only peak in
group $g_0$ that occurred in both 2015 and 2016. It is a variability
in its own right as plain beat frequencies do not appear in power
spectra.  Figure \ref{28LCf4} illustrates the difference well: The
rise in amplitude of $f_{\Delta32}$ at the end of the observations in
2016 sets in rather suddenly and after an extended period of
quiescence.  Similarly rapid rises and drops in the amplitude of
$\Delta$ frequencies have also been seen in \object{25\,$\psi^1$ Ori}
\citep{2016arXiv161101113B}.  A simple beating of two frequencies
looks different.

During its maximum, the amplitude of $f_{\Delta32}$ reaches a PTV
value of 100\,mmag.  This is well outside the domain of nonradial
pulsations in other B-type stars and exceeds the (time-independent,
see Fig.\,\ref{VarAmp1516}) amplitude sum of its parent frequencies
$f_2$ and $f_3$ by a factor of 4-5.  Other stars with $\Delta$
amplitudes above the sum of the parent amplitudes include
\object{$\eta$\,Cen} \citepalias{2016A&A...588A..56B},
\object{10\,CMa}, and \object{25\,$\psi^1$\,Ori}
\citep{2016arXiv161101113B}.  Moreover, the variability with
$f_{\Delta32}$ is highly non-sinusoidal with the range above the
inflection points of the light curve being much larger than that below
(Fig.\,\ref{28LCf4}). This asymmetry and the large amplitude can be
understood as near-circumstellar matter emitting extra light
\citep{2012ApJ...756..156H, 2016A&A...588A..56B}.  Such
large-amplitude asymmetries have also been observed by BRITE
Constellation in 25\,$\psi^1$\,Ori \citep{2016arXiv161101113B}.
Because of the EER effects, it is not possible to infer the real
photospheric amplitude associated with $f_{\Delta32}$, and the above
comparison of $\Delta$-frequency amplitude and parent $g$-mode
amplitudes has no energetic meaning.

Figure \ref{28LCf4} reveals another important detail: In the
$f_{\Delta32}$ cycle around mJD\,630, several measurements appear up
to $\sim$100\,mmag below the upper envelope of the light curve.  They
reach well below the mean brightness so that there is a temporary
deficit of light.  These fadings should not be instrumental because
both UBr and BTr independently recorded a similar, slightly weaker,
event around mJD\,198 (Fig.\,\ref{28LCf4}).  The reality of this
dimming is undoubtable.  But the absence of simultaneous observations
at all other times makes it impossible to assess the reality of any of
the other, mostly much sparser, groups of bright and faint points.
However, Figs.\,\ref{28LCf4scat2015} and \ref{28LCf4scat2016} do point
to increased scatter around extrema of the $f_{\Delta32}$ light curve.

Comparable rapid switching between increased and attenuated brightness
during phases of maximal $\Delta$ amplitude has been observed by BRITE
in 25\,$\psi^1$\,Ori \citep{2016arXiv161101113B}.  This star has a
spectral type and $v$\,sin\,$i$ similar to 28\,Cyg so that it is also
viewed from a similar perspective.  One possible interpretation (cf.\
Sect.\,\ref{ansatz}) is, therefore, that ejecta span some range in
stellar latitude.  The emission by the lifted-up matter is permanent
(but not constant) whereas the only temporary removal of light implies
strong variability in the amount of matter along the line of sight to
the photosphere.  This corroborates the conclusion that at the end of
the BRITE observations in 2016 28\,Cyg suffered a (small) series of
mass ejections.  That is, these brightenings probably were real
outbursts.

With this behaviour of \object{28\,Cyg} (and
\object{25\,$\psi^1$\,Ori}) there are three ways how a $\Delta$
frequency can sculpt mass-loss from Be stars:

\noindent
1.) Spectroscopy of \object{$\mu$\,Cen} \citep{1998ASPC..135..343R} has
shown that major enhancements of the H$\alpha$ line emission happen
every time when the vectorial sum of at least two out of three
specific NRP modes exceeds a threshold in amplitude.

\noindent
2.) In \object{$\eta$\,Cen} \citepalias{2016A&A...588A..56B}, the mass
loss is basically permanent as is suggested by the continuous presence
of the {\v S}tefl frequency in emission lines, which seems to be the
response by the disc to mass transfer.  The latter is continually
modulated by $f_{\Delta32}$ with minor apparently stochastic
deviations superimposed.

\noindent
3.) The same may be at work in 28\,Cyg but at a lower level
(Fig.\,\ref{28LCf4scat2015}, Sect.\,\ref{circumstellar}).  Much more
prominent are single events spaced with the $\Delta$ period.  But the
series of such events starts suddenly and, by inference from
25\,$\psi^1$\,Ori, also declines quickly (the observations in 2016 of
28\,Cyg were terminated before the end of the active phase because the
Sun was too close).  This confirms the report by
\citet{2000ASPC..214..232T} that they saw only one H$\alpha$
line-emission outburst and could rule out repetitions with
$f_{\Delta32}$\,=\,0.05\,c/d during their 100-d and 65-d observing
windows.

It remains to be seen whether these three categories are genuine or
the artifact of incomplete temporal sampling.  There is no explanation
of the timing of the brightenings in 28\,Cyg.  They may involve a
third clock (in addition to $f_2$ and $f_3$).

\subsection{$f_1$ as a {\v S}tefl frequency and other circumstellar 
variabilities}
\label{circumstellar}

\citetalias{2016A&A...588A..56B} called the high-amplitude 1.56-c/d
frequency of \object{$\eta$\,Cen} a {\v S}tefl frequency after the
discoverer of this kind of variability \citep{1998ASPC..135..348S}.
In \object{$\eta$\,Cen}, the {\v S}tefl frequency was
spectroscopically detected \citep{2003A&A...411..229R}.  Typical {\v
  S}tefl frequencies occur during outbursts, are best seen in emission
lines and in absorption lines formed high in the atmosphere, and - for
unknwon or accidental reasons - appear to have values $\sim$10\%
below the main spectroscopic frequency.  Condensed to the capabilities
of photometry, this is a good initial description of $f_1$ in
\object{28\,Cyg}.

In $\eta$\,Cen, both amplitude and frequency of the {\v S}tefl
frequency are strongly variable and appear modulated with the $\Delta$
frequency \citepalias{2016A&A...588A..56B}.  In 28\,Cyg, the
variations are also large but there is no regularity in the frequency,
only a major perturbation just before the series of brightenings in
2016 (Figs.\,\ref{timefreq} and \ref{VarFrq13800}).  This large
frequency jitter is also reminiscent of Achernar, in which
\citet{2011MNRAS.411..162G} meticulously documented phase shifts of
one of the two detected frequencies.

The maxima of the brightenings are roughly matched by strong amplitude
maxima of $f_1$ (compare Figs.\,\ref{28LCf4}, \ref{timefreq}, and
\ref{VarAmp1516}).  This is compatible with the spectroscopically
established notion of \citetalias{2016A&A...588A..56B}: When an
outburst/brightening occurs, the distribution of matter in the inner
disc is not yet homogeneous \citep[][]{1988A&A...198..211B,
  1991ESOC...36..185H}.  The level of inhomogeneity reveals itself by
the amplitude of the {\v S}tefl frequency.  As in \object{$\eta$\,Cen}
(and other Be stars observed by BRITE), the $f_1$ is embedded in a
frequency group ($g_1$; Figs.\,\ref{28powerB}, \ref{PS16lohi}, and
\ref{28powerNarrowS}) the strength of which is crudely related to that
of the $f_1$.  In 28\,Cyg, a fair part of the neighbourhood of $f_1$
changes during a major brightening (Fig.\,\ref{PS16lohi}).  Both
observations probably strengthen the notion that these apparent
companion frequencies are circumstellar.  On all these the grounds,
the classification of $f_1$ as the {\v S}tefl frequency of 28\,Cyg
appears safe.

At 9\,M$_\odot$, the model of \citet{2012ApJ...756..156H} should be a
good match of a B2 star like 28\,Cyg
\citep[cf.][]{1988BAICz..39..329H}.  At the assumed typical fractional
critical rotation rate of 80\% \citep{2005A&A...440..305F,
  2012A&A...538A.110M, 2013A&ARv..21...69R} its equatorial radius is
increased from 5.7\,R$_\odot$ to 6.5\,R$_\odot$.  The orbital radius
of matter circling this star with the {\v S}tefl frequency, 1.4\,c/d,
amounts to 6.9\,R$_\odot$.  Comparison of the two radii not only
supports the exophotospheric interpretation of the {\v S}tefl
frequency but is also consistent with the model prediction that, at
optical wavelengths, the emission from exophotospheric matter arises
from a region measuring less than about two equatorial radii.

In spite of all these details, the net effect of the brightenings in
2016 remains uncertain.  The event reported by
\citet{2000ASPC..214..232T} impacted a fair part of the disc.  But the
cadence of the spectroscopy did not trace accompanying star-disc
connections.  The BRITE photometry has the necessary sampling but
cannot equally clearly distinguish between photospheric and
exophotospheric variability.  BRITE photometry and {\sc Heros}
spectroscopy can be reconciled if the 1998 drop in H$\alpha$ emission
strength was caused by a short increase of the continuum flux similar
to that recorded by BRITE in 2016.  This would be in good agreement
also with the models of \cite{2012ApJ...756..156H}.

The contemporaneous BeSS spectroscopy (Fig.\,\ref{VtoR}) exhibits no
direct counterpart of the photometric brightenings.  A most sensitive
indicator could be the wings of the H$\alpha$ line emission.  The
effects of the high orbital velocity of, and Thomson scattering by,
inner disc matter should both strengthen the wings
\citep{1986BIEBe..13....5B}, but only in dense discs.  At the time of
the BeSS spectroscopy in 2016 the disc of 28\,Cyg was only moderately
developed.  This probably explains why enhanced H$\alpha$ wings wer
not detected (Fig.\,\ref{BeSSperspHalpha}) in the wake of the 
brightening around mJD\,630.

Figure\,\ref{VtoR} could be read as follows: At some moment, fresh
matter was injected into the disc.  With the help of viscosity, part
of it drifted outward but most of it lost angular momentum and fell
back to the star.  Because the disc is optically thick in H$\alpha$,
the equivalent width does not change immediately after the outburst
but only starts to increase when the outward drifting matter reaches a
domain that was not optically thick before.  Since this region is
farther away from the star and the disc is Keplerian, the separation
of the emission peaks decreases.

\begin{figure}
\includegraphics[width=9cm,angle=0]{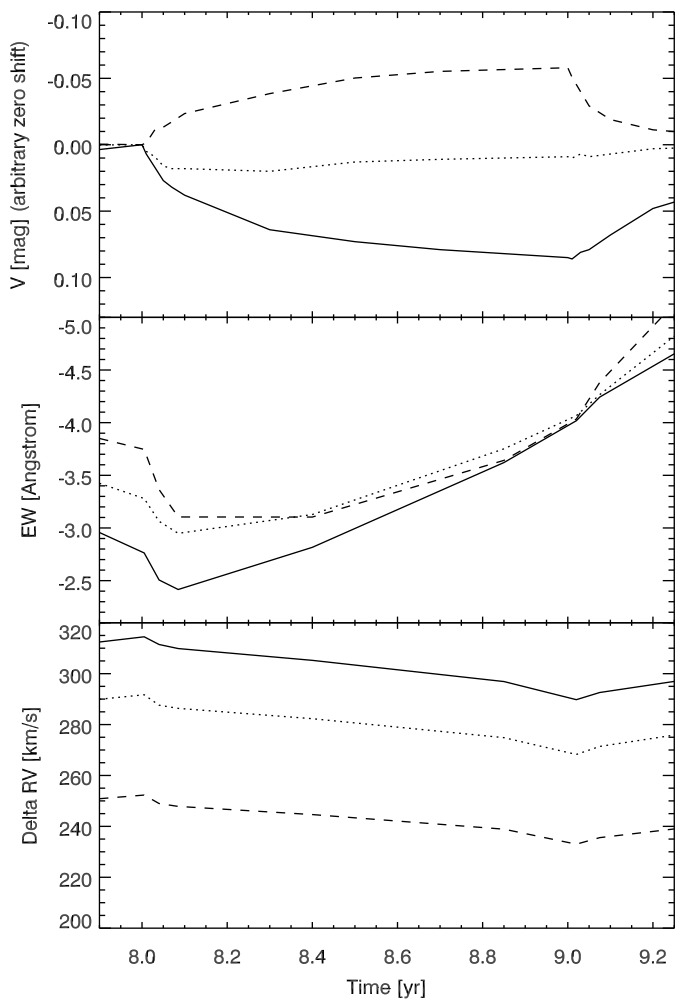}
\caption{Temporal evolution of the $V$-band brightness (top),
  H$\alpha$ line equivalent width (middle), and peak separation
  (bottom) computed with the periodic model of \citet[][their
  Fig.\,7]{2012ApJ...756..156H} consisting of a series of one-year
  disk build-up phases followed by one-year dissipation phases. Solid,
  dotted, and dashed lines represent inclination angles $80^\circ$,
  $70^\circ$, and $60^\circ$, respectively.  Note the difference in
  sign between the $V$ magnitude changes at $80^\circ$ and $60^\circ$.
  The outburst begins at $t = 8$\,yr and ends at $t = 9$\,yr.}
\label{VDD} 
\end{figure}

This description is consistent with VDD hydrodynamic simulations
coupled with radiative transfer calculations
\citep{2012ApJ...756..156H}.  Fig.\,\ref{VDD} illustrates a periodic
model \citep[see Fig.\,7 of][]{2012ApJ...756..156H} that alternates
between one year of disc build-up at a fixed disc-feeding rate with
one year of disc dissipation when the mass loss is shut off.  At
$t=8$\,yr, the outburst begins, replenishing the remnant of the disc
from the previous two-year cycle with fresh material. Initially the
line equivalent width drops slightly, as a result of the increased EER
continuum emission, but then raises steadily and continues to grow
past the cessation of the disk feeding at $t=9$\,yr owing to the quick
decrease in the EER continuum emission. As observed in
\object{28\,Cyg} (Fig.\, \ref{VtoR}), the peak separation
anticorrelates with line emission, but this is no longer true during
disc dissipation.

Although $f_1$ exhibits large amplitude variations, it never
disappears completely.  If {\v S}tefl frequencies are indeed close
tracers of star-to-disc mass transfer, the implication is that this
mass transfer was variable but permanent during the BRITE observations
of 28\,Cyg.  Such continuous, albeit variable, process may make larger
contributions to the regular disc replenishment needed to overcome the
demolishing effect of the viscosity than outbursts do.  This is
another important commonality with $\eta$ Cen
\citepalias{2016A&A...588A..56B}.  It seems supported also by the
persistence, for five months, of fairly strong wings of the not so
strong H$\alpha$ line emission (Fig.\,\ref{BeSSperspHalpha}).  Even in
a late-type Be star, whose disc may additionally be truncated by a
companion star, the H$\alpha$ line emission forms over many stellar
radii \citep{2015A&A...584A..85K}.  Therefore, the inner disc volume,
when well filled, acts as a buffer that is insensitive to minor
variations of the mass inflow.

The amplitude of $f_1$ was also increased at the end of the
observations in 2015 (Figs.\,\ref{timefreq} and \ref{VarAmp1516}) when
there was no hint of a major brightening (Fig.\,\ref{28LCf4}).  This
might indicate that the steady mass-loss component can also lead to
inhomogeneous mass distributions close to the star (unless due to
intrinsic local instabilities), or the outburst was not strong enough
to develop a disc.

Unless winds of Be stars originate from the circumstellar discs
\citep{2013A&ARv..21...69R}, in which case a direct link to pulsation
is not expected, it would be interesting to repeat the simultaneous UV
and optical spectroscopy by \citet{2000ASPC..214..375P} to identify
which of $f_1$, $f_2$, and $f_3$, if any, correlates with UV flux and
wind.  To date, in no Be star, including 28\,Cyg, has precision
photometry or spectroscopy found mass loss clearly modulated with the
frequency of a single NRP mode.  Instead, timescales 20-50 times
longer (occurring as $\Delta$ frequencies) have been inferred.  It
would be useful to extend UV observations to this range and also to
check for the distribution of the variability between sub- and
super-escape velocities.  The sampling should permit detection of the
circumstellar $f_1$.

\section{Conclusions and outlook}
\label{conclusions} 

In the photometric variability of early-type Be stars some incipient
commonalities emerge as a set of frequencies that follow a common
pattern whose numerical values form the specific fingerprint of every
star.  The top-level variability of 28\,Cygni is structurally the same
as that of the very similar star $\eta$\,Cen
\citepalias{2016A&A...588A..56B}:

\noindent
(i) Most frequencies are concentrated in groups
(Fig.\,\ref{28powerB}).

\noindent 
(ii) Four frequencies dominate the short- and medium-term
variability. They fall into three different categories:

\noindent
(iia) Two spectroscopically confirmed $g$ modes \citep[$f_2$ and 
$f_3$,][]{2000ASPC..214..232T}.  

\noindent
(iib) One {\v S}tefl frequency ($f_1$), which originates from the
star-disc transition zone and is about 10\% lower in value than $f_2$
and $f_3$.  It may occasionally become undetecable, and it can
drastically increase around brightenings but does not contribute to
them (Fig.\,\ref{VarAmp1516}).  Its frequency is much less stable than
the frequencies of the $g$ modes (Figs.\,\ref{timefreq} and
\ref{VarFrq13800}).  But its avarage value is roughly constant at all
times.

\noindent
(iic) One $\Delta$ frequency ($f_{\Delta32}$) of the two $g$ modes.
  It has the largest amplitude.  During amplitude maxima, star-to-disc
  mass loss seems increased as indicated by the enhancement of the {\v
    S}tefl variability of emission lines in $\eta$\,Cen (Paper I; for 28
  Cyg, see Figs.\,\ref{28LCf4}, \ref{28LCf4scat2015}, and
  \ref{28LCf4scat2016}).  This process may also be persistently
  modulated with the $\Delta$ frequency.

\noindent
At this level, $\eta$\,Cen and 28\,Cyg only differ in the numerical
values of the parameters quantifying the frequencies and their
amplitudes.  Both stars also share the long-term persistence of their
basic variability patterns, including considerable amplitude
variations especially of $f_1$ and $f_{\Delta32}$.

The existence of a second star like $\eta$\,Cen lends considerable
support to the description of the variability given in
\citetalias{2016A&A...588A..56B}.  An inner `engine' clocked by the
$g$ modes and powered by the $\Delta$ variability lifts matter above
the photosphere.  In the inner disc, a viscosity-ruled outer engine
sorts the matter delivered to it into two categories, namely super-
and sub-Keplerian specific angular momentum.  Part of this process is
observed as {\v S}tefl frequency, the amplitude of which increases
with the total amount of inner-disc matter and its deviation from
circular symmetry.  It is not yet possible to guess whether or not
$\eta$\,Cen and 28\,Cyg are representative of a significant fraction
of all early-type Be stars.

\citet[][MOST]{2005ApJ...635L..77W},
\citet[][CoRoT]{2009A&A...506...95H},
\citet[][CoRoT]{2012A&A...546A..47N},
\citet[][Kepler]{2015MNRAS.450.3015K}, and others have argued that the
myriads of frequencies in Be stars are due to pulsations.  With so
much support, this conclusion is not to be dismissed lightly.
However, Papers I and II of this series have suggested on several
grounds that a circumstellar origin may be a fairly reasonable
alternative for a good part of this variability.  The coming and going
of many features in the frequency spectrum also of \object{28\,Cyg}
(Sect.\,\ref{ephemeral} and Fig.\,\ref{timefreq}) may also point at
the {\v S}tefl frequency as the tip of the iceberg.  

Therefore, it makes sense to search for the key to the understanding
of the Be phenomenon in the small set of anchor frequencies instead of
transient variations.  Only for them is there any plausible direct
observed link to mass loss (Figs.\,\ref{28LCf4}, \ref{28LCf4scat2015},
and \ref{28LCf4scat2016}).  Ultimately, this requires that undulations
of the short-term variability can also explain disc life cycles of up
to a decade inferred from the H$\alpha$ emission-line strength (taken
as a proxy of the total amount of matter in the circumstellar disc).
In 28\,Cyg, the anchor frequencies have persisted for several decades
which is long enough to let them enter into the explanation.

The simplest extrapolation in timescale to nearly a decade (0.1 c/yr)
would be a combination of modes with $\Delta$ frequencies of order 0.1
c/yr.  However, viscosity will not let single big outbursts every
10\,yr create photometric high states lasting for several years.  In
the VDD paradigm, a photometric high state of a disc could result from
outbursts with a frequency and effectiveness sufficient to sustain a
massive disc.  The lowest level would be formed by mini-outbursts with
single $\Delta$ frequencies.  A second clock could come into play for
outbursts perhaps associated with the brightening in \object{28\,Cyg}
around mJD\,630.  At least a third level would be required to regulate
the first two processes and cause cycles of a few years length.  The
example of the brightening around mJD\,630 illustrates that the
amplitude amplification can be a highly nonlinear process, either in
true NRP amplitudes or their effect on mass loss or both.

One way of achieving slow clocking could be 'hierarchical' $\Delta$
frequencies, i.e., combinations of $\Delta$ frequencies.  In
\object{28\,Cyg}, they could be $f_{\Delta32}$ and, if real,
0.00995\,c/d (Sect.\,\ref{ephemeral}).  In their most basic form, they
might explain seemingly particularly simple cases as depicted in
Fig.\,5 of \citet{2002AJ....124.2039K} in which long-lasting bright
states cyclically repeat on two similar timescales of order
$500-1000$\,d.

Irrespective of how several NRP wave patterns combine to $\Delta$
frequencies, it seems plausible that for best coupling efficiency
their angular structures must be the same, i.e., they should agree in
their $\ell$ and $m$ indices.  In $\mu$ Cen
\citep{1998A&A...336..177R} and 28\,Cyg \citep{2000ASPC..214..232T}
the available spectroscopic evidence supports this expectation.  The
work of \citet{2003A&A...411..229R} has furthermore demonstrated that
$\ell$ = $-m$ = 2 modes dominate the spectral variability of many Be
stars.  If equality of the mode indices is a requirement for
mass-loss-modulating difference frequencies, this could be an
important filter meaning that not any two arbitrary NRP modes can
combine to cause mass loss.  Indications of mass lift-ups occurring
over some range in stellar latitude may place first constraints on the
pulsational velocity field ($\ell$ and $m$).  However, spectroscopic
searches for $\Delta$ frequencies could be more effective in
determining the geometry of the variability.  From models, it would
be attractive to learn whether energy leakage from pulsations to mass
loss can establish another effective filter of the NRP mode spectrum.

A robust distinction between stellar and circumstellar phenomena is
critical for the identification of frequencies for asteroseismology.
High-cadence long-term space photometry has the potential of achieving
this.  For the understanding of the Be phenomenon at large it is
comforting that with the VDD model a tool is available that permits
observed circumstellar variabilities to be cast into coherent
sequences of dynamical states.

\begin{acknowledgements} The authors thank the BRITE operations staff
  for their untiring efforts to deliver data of the quality that
  enabled this investigation.  They are also very grateful to the many
  amateur astronomers who, in response to a specific request, took
  numerous H$\alpha$ spectra and shared them through BeSS.  This
  research has made use of the SIMBAD database, operated at CDS,
  Strasbourg, France.  This research has made use of NASA's
  Astrophysics Data System.  This work has made use of the BeSS
  database, operated at LESIA, Observatoire de Meudon, France:
  http://basebe.obspm.fr.  ACC acknowledges the support from CNPq
  (grant 307594/2015-7) and FAPESP (grant 2015/17967-7).  DP
  acknowledges financial support from Conselho Nacional de
  Desenvolvimento Cient\'ifico e Tecnol\'ogico (CNPq-MCTIC Brazil;
  grant \mbox{300235/2017-8}). M.R.G. acknowledges the support from
  CAPES PROEX Programa Astronomia.  GH thanks the Polish NCN for
  support (grant 2015/18/A/ST9/00578).  AFJM is grateful for financial
  aid from NSERC (Canada) and FRQNT (Quebec).  APi acknowledges
  support from the Polish NCN grant no. 2016/21/B/ST9/01126.  APo
  acknowledges support through NCN grant No.\ 2013/11/N/ST6/03051.
  GAW acknowledges Discovery Grant support from the Natural Sciences
  and Engineering Research Council (NSERC) of Canada.  The Polish
  contribution to the BRITE project is supported by Polish Ministry of
  Science and Higher Education, and the Polish National Science Center
  (NCN, grant 2011/01/M/ST9/05914).
\end{acknowledgements}

\bibliography{dbaade}

\end{document}